\pdfoutput=1
\documentclass[usenatbib,useAMS,letterpaper]{mn2e}
\usepackage[pdftex]{graphicx}
\usepackage{epstopdf}
\usepackage{ctable}
\usepackage{url}
\usepackage{times}

\newcommand{\lsun}{~\mathrm{L_{\odot}}}
\newcommand{\lbol}{L_{\rm bol}}
\newcommand{\lir}{L_{\rm IR}}

\newcommand{\msun}{~\mathrm{M_{\odot}}}
\newcommand{\msunperyr}{~\mathrm{M_{\odot} {\rm ~yr}^{-1}}}

\newcommand{\mgas}{M_{\rm gas}}
\newcommand{\mstar}{M_{\star}}
\newcommand{\mdust}{M_\mathrm{d}}
\newcommand{\tdust}{T_\mathrm{d}}
\newcommand{\tdustot}{T_\mathrm{d,OT}}
\newcommand{\tc}{T_\mathrm{c}}

\newcommand{\dbh}{d_\mathrm{BH}}
\newcommand{\lfuv}{L_{\rm FUV}}

\newcommand{\sunrise}{\textsc{Sunrise}~}
\newcommand{\gadget}{\textsc{Gadget}~}
\newcommand{\gadgettwo}{\textsc{Gadget-2}~}

\newcommand{\arepo}{\textsc{Arepo}~}

\newcommand\plotone[1]
 {\centering \leavevmode \includegraphics[width={0.99\columnwidth}]{#1}}

 \newcommand\plottwo[2]{{%
 \typeout{Plottwo included the files #1 #2}
 \centering
\includegraphics[width={0.99\columnwidth}]{#1}%
\\
\includegraphics[width={0.99\columnwidth}]{#2}%
}}%
\newcommand{\acknowledgments}{\begin{small}\section*{Acknowledgments}\end{small}}

\title[Star formation modes \& the SMG bimodality]{How to distinguish starbursts and quiescently star-forming galaxies:
The `bimodal' submillimetre galaxy population as a case study}
\author[Hayward et al.]{
\parbox[t]{\textwidth}{
Christopher C. Hayward$^{1,2}$\thanks{E-mail: Christopher.Hayward@h-its.org},
Patrik Jonsson$^2$, Du\v{s}an Kere\v{s}$^{3,4}$\thanks{Hubble Fellow},
Benjamin Magnelli$^{5}$,  Lars Hernquist$^{2}$, \& T.~J. Cox$^{6}$
}
\vspace*{6pt} \\
$^{1}$Heidelberger Institut f\"ur Theoretische Studien, Schloss--Wolfsbrunnenweg 35, 69118 Heidelberg, Germany \\
$^2$Harvard--Smithsonian Center for Astrophysics, 60 Garden Street, Cambridge, MA 02138, USA \\
$^3$Department of Physics, Center for Astrophysics and Space Science, University of California at San Diego, 9500 Gilman Drive, La Jolla, CA 92093, USA \\
$^4$Department of Astronomy and Theoretical Astrophysics Center, University of California Berkeley, Berkeley, CA 94720, USA \\
$^5$Max--Planck--Institut f\"ur Extraterrestrische Physik, Postfach 1312, 85741 Garching, Germany \\
$^6$Carnegie Observatories, 813 Santa Barbara Street, Pasadena, CA 91101, USA}

\begin{document}

\date{Accepted 2012 May 4. Received 2012 April 23; in original form 2012 March 5}

\pagerange{\pageref{firstpage}--\pageref{lastpage}} \pubyear{2012}

\maketitle

\label{firstpage}

\begin{abstract}
In recent work \citep{Hayward:2011smg_selection}
we have suggested that the high-redshift ($z \sim 2-4$) bright submillimetre galaxy (SMG) population is heterogeneous, with major mergers contributing
both at early stages, where quiescently star-forming discs are blended into one submm source (`galaxy-pair SMGs'), and late stages,
where mutual tidal torques drive gas inflows and cause strong starbursts.
Here we combine hydrodynamic simulations of major mergers with 3-D dust radiative transfer calculations to determine observational diagnostics that can distinguish
between quiescently star-forming SMGs and starburst SMGs via integrated data alone.
We fit the far-IR SEDs of the simulated galaxies with the optically thin single-temperature modified blackbody, the full form of the single-temperature
modified blackbody, and a power-law temperature-distribution model. The effective dust temperature, $\tdust$, and power-law index of the dust emissivity in the far-IR, $\beta$, derived
can significantly depend on the fitting form used, and the intrinsic $\beta$ of the dust is not recovered.
However, for all forms used here, there is $\tdust$ above which almost all simulated galaxies are starbursts, so a $\tdust$ cut is very effective at selecting starbursts.
Simulated merger-induced starbursts also have higher $\lir/\mgas$ and $\lir/\lfuv$ than quiescently star-forming galaxies
and lie above the star formation rate--stellar mass relation. These diagnostics can be used to test our claim that the SMG population
is heterogeneous and to observationally determine what star formation mode dominates a given galaxy population.
We comment on applicability of these diagnostics to ULIRGs that would not be selected as SMGs. These `hot-dust ULIRGs' are typically
starburst galaxies lower in mass than SMGs, but they can also simply be SMGs observed from a different viewing angle.
\end{abstract}


\section{Introduction}

\subsection{The two modes of star formation} \label{S:SF_modes}

Star formation is one of the fundamental processes driving galaxy formation: it depletes the gas content of galaxies, enriches the
interstellar medium (ISM) with metals, and deposits
energy and momentum via supernovae, stellar winds, and radiation pressure. Furthermore, the light emitted by stars encodes much information about the current
physical properties of a galaxy and the galaxy's formation history. Thus understanding star formation is crucial for understanding galaxy formation and evolution.

An important step toward understanding the star formation processes that built up galaxies across cosmic time is determining where and when most stars are formed,
be it in disc galaxies or in starbursts triggered by, e.g., galaxy mergers, which are short-lived but can dramatically alter a galaxy's properties.
An increasing amount of observational evidence supports the notion that there are two modes of star formation \citep[e.g.,][]{Wuyts:2011a,Wuyts:2011b,Elbaz:2011,Rodighiero:2011,
Nordon:2012}, typically referred to as quiescently star-forming or quiescent\footnote{Confusingly,
the term `quiescent' is also used to refer to galaxies that have essentially no ongoing star formation; here the term `quiescent' always means `quiescently star-forming'.}
(that occurring
in normal disc galaxies) and starburst (found in unstable discs and merging galaxies at first passage and coalescence, though whether a starburst is induced in
the latter depends on factors such as gas content, orbit, and mass ratio of the progenitors; e.g., \citealt{Cox:2008}).
One clear difference is that gas depletion time-scales
of starbursts are significantly lower than those of quiescently star-forming galaxies. Star formation in starbursts tends to be dominated by the nuclear
regions, whereas quiescent star formation is more extended. Additionally, starbursts may obey a different
global Kennicutt--Schmidt (KS) relation \citep{Kennicutt:1998,Schmidt:1959} than quiescently star-forming disc galaxies:
\citet{Daddi:2010} and \citet{Genzel:2010} argue that the normalisation of the KS relation for starbursts is $\sim4-10\times$
greater than that for quiescently star-forming discs. However, this conclusion strongly depends on the bimodality of the values adopted for the CO--H$_2$ conversion
factor, so the data in fact may be consistent with a single KS law normalisation \citep{Narayanan:2011X_CO_II}.
The ratio of infrared (IR) luminosity to molecular gas mass, which is a measure of the global star formation efficiency of the system, is larger in starbursts than in
quiescently star-forming galaxies, though the magnitude of the difference is also sensitive to the CO--H$_2$ conversion factor.
Furthermore, the relationship between SFR and dust mass may also show a bimodal behaviour \citep{da_Cunha:2010}.

At a given redshift, most galaxies lie on a tight relation between SFR and stellar mass
\citep[$M_{\star}$;][]{Noeske:2007b,Noeske:2007a,Daddi:2007,Rodighiero:2010,Karim:2011}.
The relation arises because star formation is supply-limited, so, on average, SFR correlates well with cosmological gas accretion rates, which are
well-correlated with halo mass \citep{Keres:2005,Keres:2009a,FG:2011}. In this picture, starbursts are transient events that cause a galaxy to move significantly above
the SFR--$M_{\star}$ relation for a short ($\sim50-100$ Myr) time. During the burst the gas is rapidly consumed, the SFR declines sharply, and the
galaxy returns to the SFR--$M_{\star}$ relation or is quenched (i.e., drops significantly below the relation), depending on
factors such as merger mass ratio, orbit, and gas fraction.

Though there is some observational support for two star formation modes, the underlying physics is not fully understood. Thus further detailed 
observations are crucial. However, it can be difficult to observationally determine which mode of star formation dominates a given galaxy population;
this complicates efforts to understand the underlying physics. This is especially a problem at high redshift because of stricter observational limitations,
and lessons learned from the local universe may not apply to high-redshift galaxies.
For example, at high redshift gas accretion rates are significantly higher than locally \citep[e.g.,][]{Keres:2005}, so gas fractions
\citep{Erb:2006,Tacconi:2006,Tacconi:2010,Daddi:2010} and star formation rates
\citep{Noeske:2007b,Noeske:2007a,Daddi:2007} of galaxies at fixed galaxy mass increase rapidly with redshift.
Consequently, even a typical star-forming galaxy at $z \sim 2$ can reach ULIRG luminosities
\citep[e.g.,][]{Daddi:2005,Daddi:2007,Hopkins:2008cosm_frame1,Hopkins:2010IR_LF,Dannerbauer:2009}.
It would thus be useful to have simple observational diagnostics that can be used to determine which mode of star formation dominates a given galaxy or galaxy population.
This is one of the goals of this paper.

\subsection{The `bimodality' of the SMG population} \label{S:bimodality}

Submillimetre galaxies (SMGs; \citealt*{Smail:1997}; \citealt{Barger:1998,Hughes:1998,Eales:1999}; see \citealt{Blain:2002} for a review)
are a class of high-redshift \citep[median $z \sim 2.3-2.6$;][]{Chapman:2005,Yun:2012} galaxies notable for their extreme luminosities \citep[bolometric luminosity
$\lbol \sim 10^{12}-10^{13} \lsun$; e.g.,][]{Kovacs:2006,Magnelli:2010,Magnelli:2012}, almost all of which is emitted in the IR. Since they seem to be powered by star formation rather than
active galactic nuclei (AGN;
\citealt{Alexander:2005,Alexander:2005b,Alexander:2008,Valiante:2007,Menendez:2007,Menendez:2009,Pope:2008MIR,
Younger:2008phys_scale,Younger:2009EGS}), they have inferred SFRs of $\sim 10^2-10^3 \msunperyr$ \citep[e.g.,][]{Coppin:2008}, much greater than those of
even the most extreme local galaxies.

It has long been known that during major mergers tidal torques drive significant amounts
of gas inward, resulting in a strong burst of star formation \citep{Hernquist:1989,Barnes:1991,Barnes:1996,Mihos:1996}.
Locally, ultra-luminous IR galaxies (ULIRGs, defined by $L_{\rm IR} > 10^{12} \lsun$)
are exclusively late-stage major mergers (e.g., \citealt*{Sanders:1996,Veilleux:2002}; \citealt*{Lonsdale:2006}), so the strong gas inflows
induced during the coalescence stage of a major merger seem
necessary to power the most luminous and rapidly star-forming galaxies.

The identification of ULIRGs with late-stage major mergers in the local universe has caused many researchers to suspect that SMGs,
some of the most IR-luminous galaxies at high redshift, are also late-stage major mergers.
There is much observational evidence that supports this picture
\citep[e.g.,][]{Ivison:2002,Ivison:2007,Ivison:2010,Chapman:2003, Neri:2003,Smail:2004,Swinbank:2004,Greve:2005,Tacconi:2006,Tacconi:2008,
Bouche:2007,Biggs:2008,Capak:2008,Younger:2008phys_scale,Younger:2010,Iono:2009,Engel:2010,Riechers:2011b,Riechers:2011a,Bussmann:2012,
Magnelli:2012}.
Furthermore, by combining hydrodynamic simulations with dust radiative transfer (RT), we have shown that simulated major mergers have
observed 850-\micron ~fluxes and typical spectral energy distributions (SEDs; \citealt{Narayanan:2010smg}; \citealt{Hayward:2011smg_selection};
but cf. \citealt{Chakrabarti:2008SMG}), stellar masses \citep{Michalowski:2011}, and CO properties \citep{Narayanan:2009} consistent with observed SMGs.
Semi-analytic models (SAMs) typically also find that merger-induced starbursts (though not necessarily major mergers, as minor-merger-induced
starbursts dominate in some models) account for the bulk of the SMG population
(\citealt{Baugh:2005,Fontanot:2007,Swinbank:2008,LoFaro:2009,Fontanot:2010,Gonzalez:2011}; but cf. \citealt{Granato:2004}).
However, \citet{Dave:2010} have claimed that there are not enough major mergers to account for the observed SMG population. Instead, they argue that a significant
fraction of the population must be massive discs fuelled by smooth accretion and minor mergers.

\begin{figure*}
  \centering
  \begin{tabular}{cccc}
    \includegraphics[width=4cm]{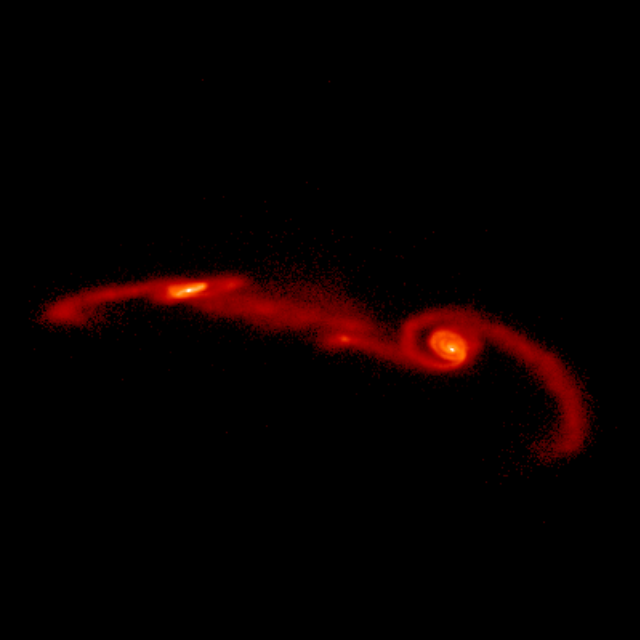} &
    \includegraphics[width=4cm]{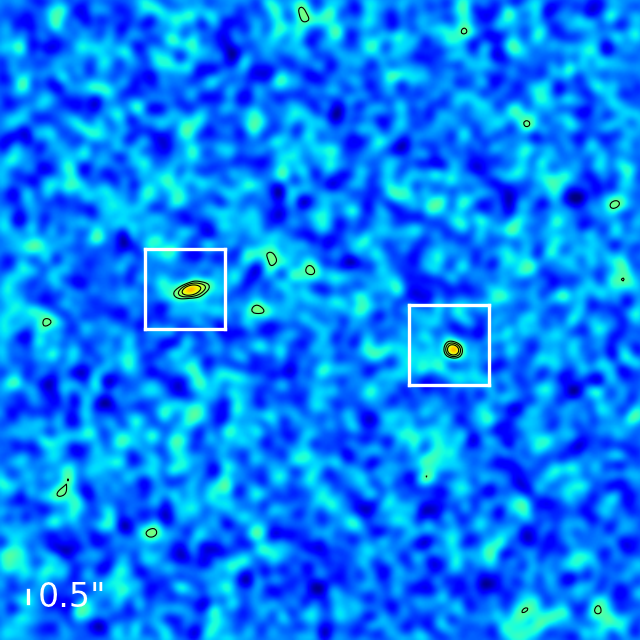} &
    \includegraphics[width=4cm]{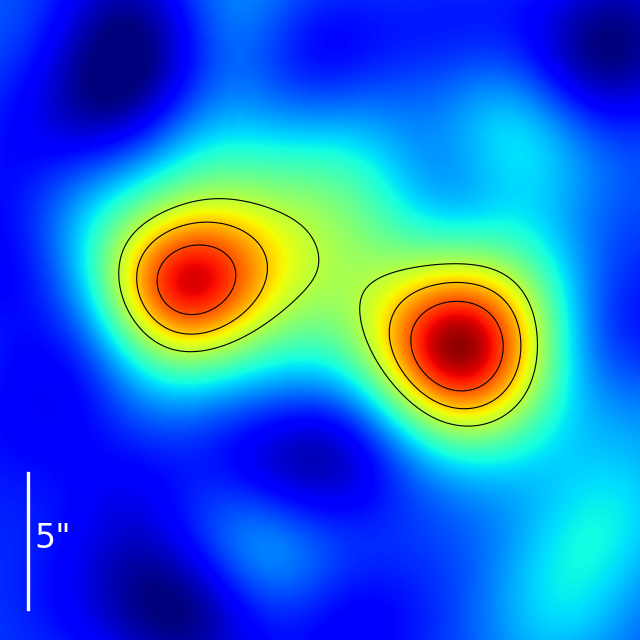} &
    \includegraphics[width=4cm]{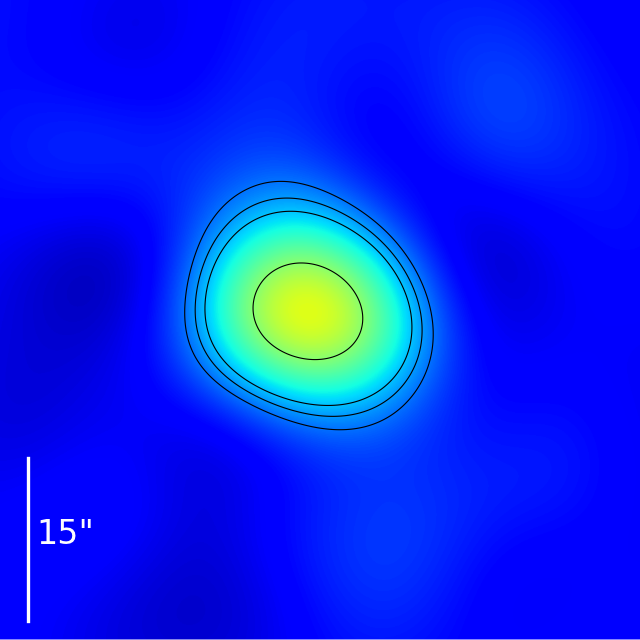} \tabularnewline
    \includegraphics[width=4cm]{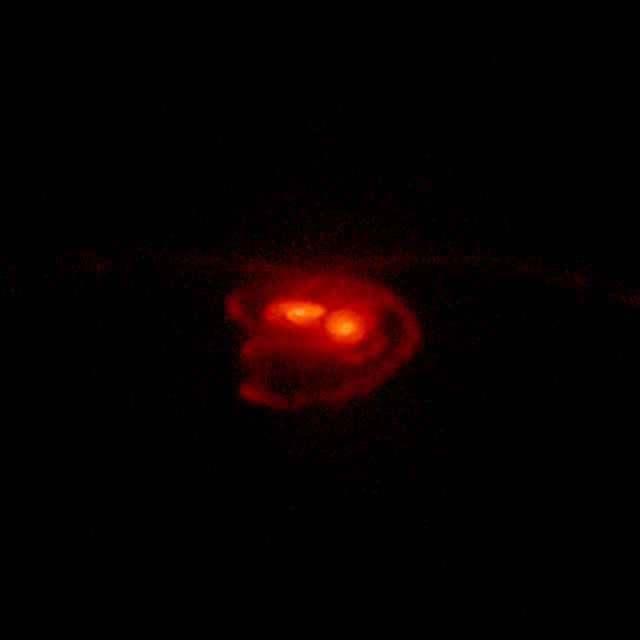} &    
    \includegraphics[width=4cm]{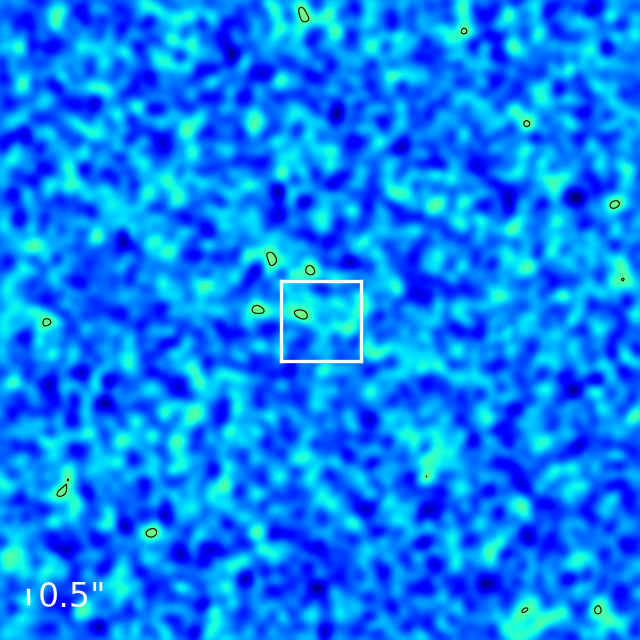} &    
    \includegraphics[width=4cm]{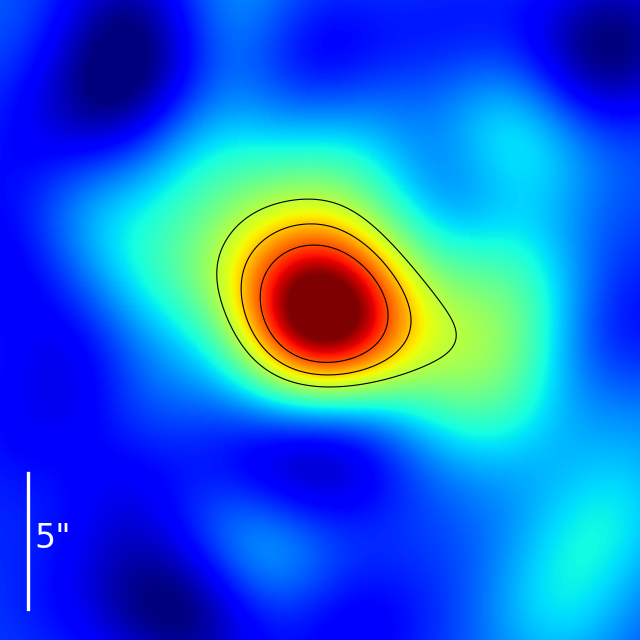} &
    \includegraphics[width=4cm]{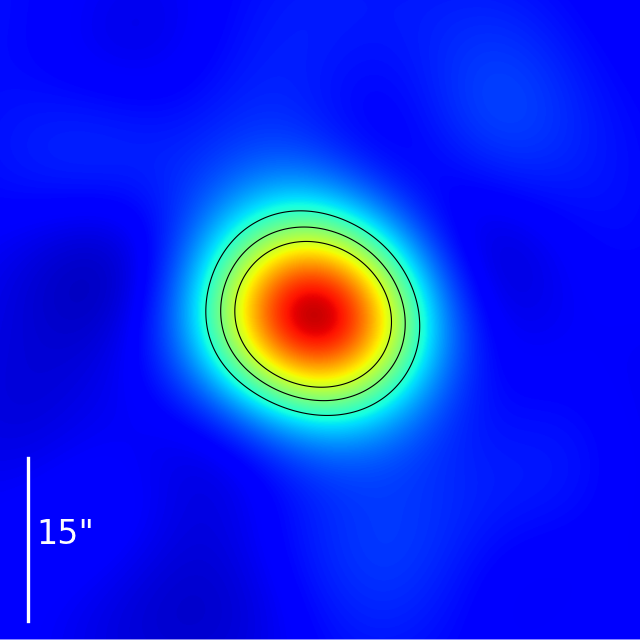} \tabularnewline
    \includegraphics[width=4cm]{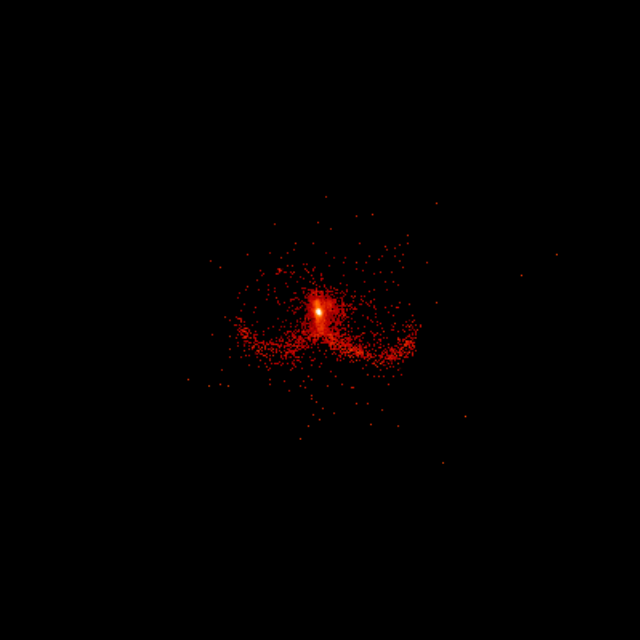} &
    \includegraphics[width=4cm]{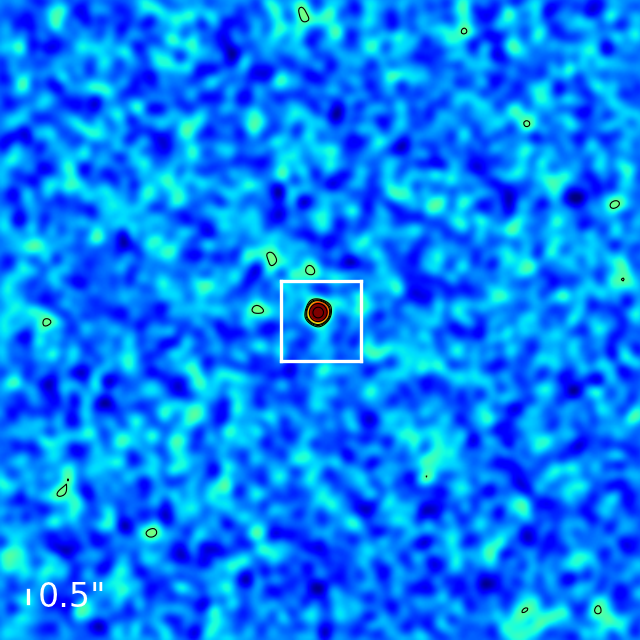} &
    \includegraphics[width=4cm]{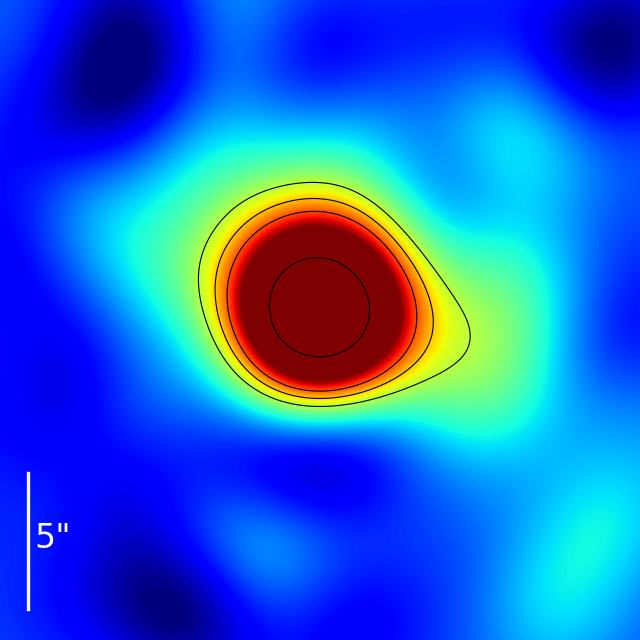} &
    \includegraphics[width=4cm]{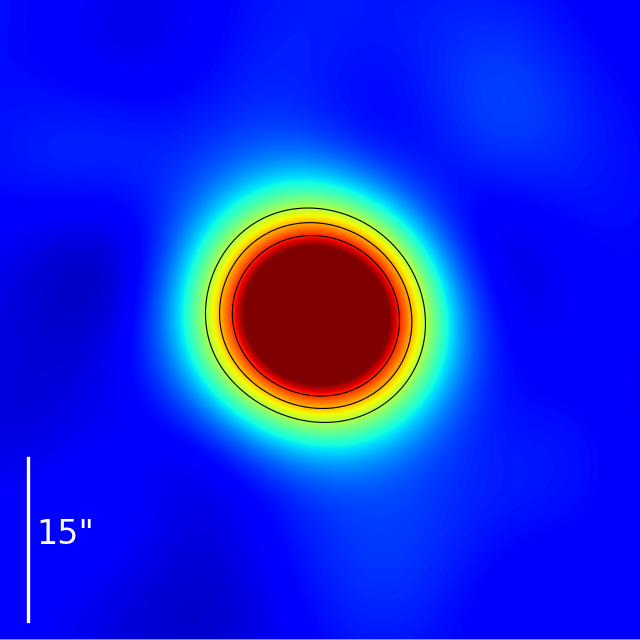} \tabularnewline
    \includegraphics[width=4cm]{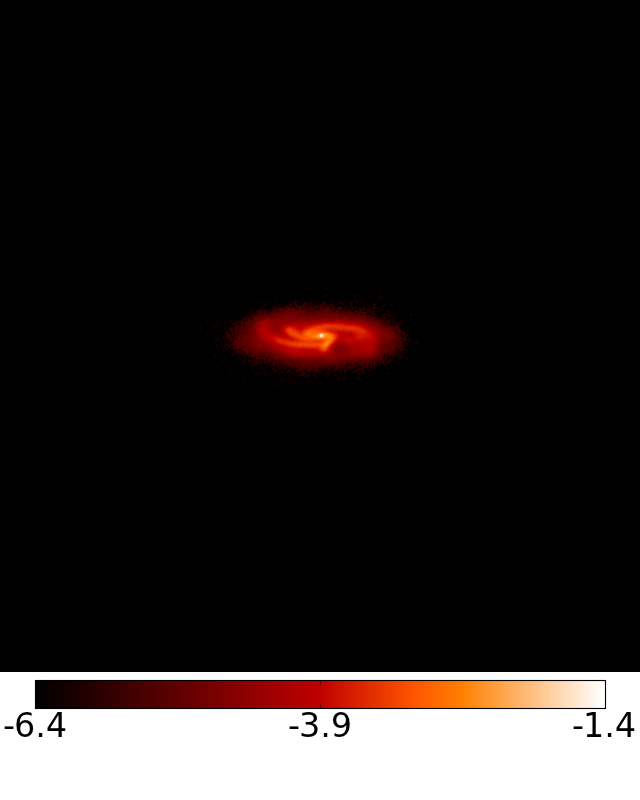} &
    \includegraphics[width=4cm]{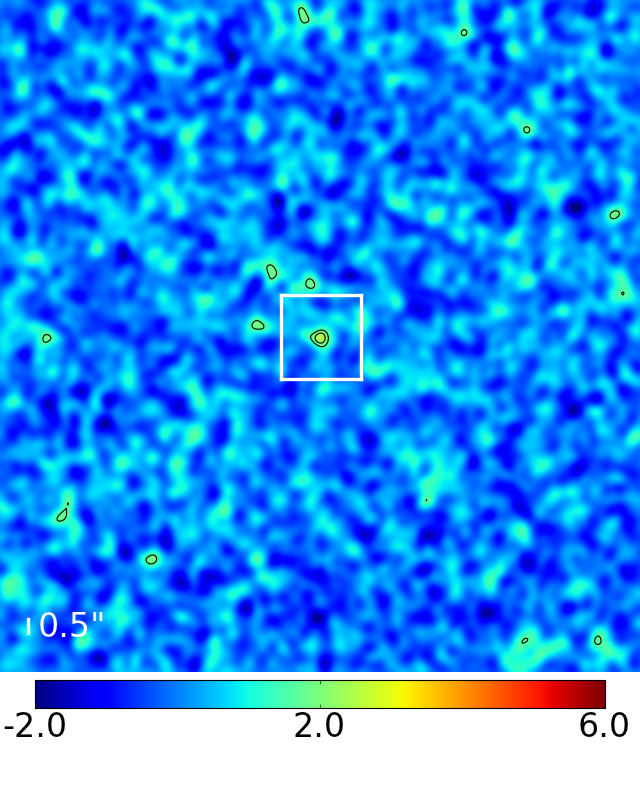} &
    \includegraphics[width=4cm]{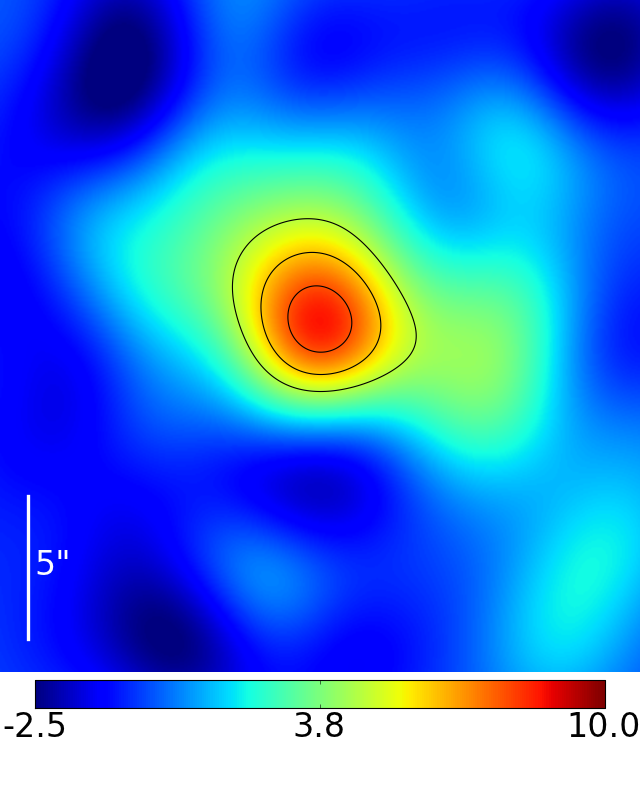} &
    \includegraphics[width=4cm]{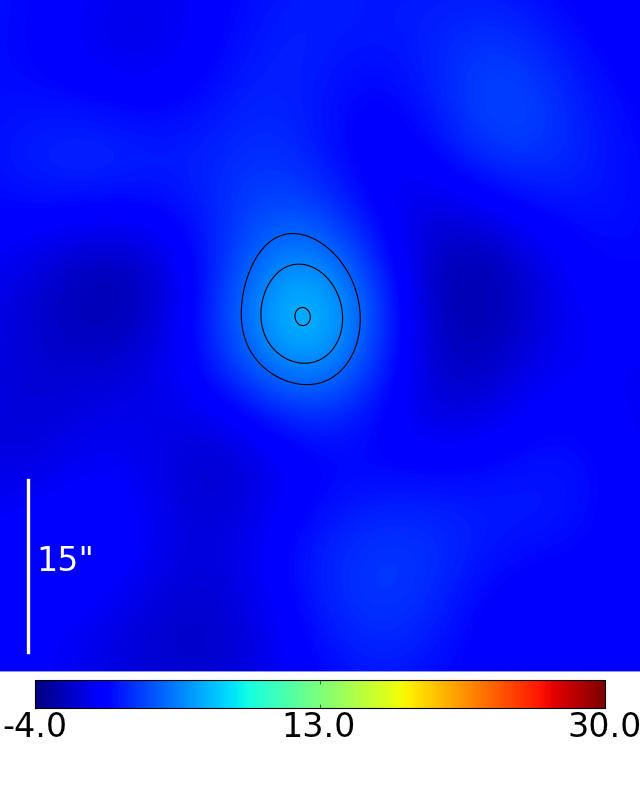} \tabularnewline
  \end{tabular}
  \caption{The first three rows show mock observed-frame 850-\micron ~continuum images (assuming $z = 2$) of the merger from H11 near apocentre (first row; time $t = 0.28$ Gyr;
  	nuclear separation $\dbh = 71 h^{-1}$ kpc; quiescently star-forming),
	during final approach (second row; $t = 0.63$ Gyr; $\dbh = 13 h^{-1}$ kpc; quiescently star-forming), and at the peak of the starburst (third row; $t = 0.7$ Gyr;
	$\dbh = 0.1 h^{-1}$ kpc; starburst). The fourth row shows the isolated disc at $t = 0.28$ Gyr (quiescently star-forming). The simulated images have been convolved with a Gaussian
	filter in order to mimic the intrinsic simulation resolution (first column; FWHM $= 0.2 h^{-1}$ kpc) and the resolution achievable with various telescopes/interferometers: The second column has FWHM
	$= 0.5$ arcsec ($3 h^{-1}$ kpc at $z = 2$), the resolution roughly characteristic of (sub)mm interferometers such as the SMA, the IRAM Plateau de Bure Interferometer, and ALMA.
	(Note we have not made any attempt to model the interferometric process.) See Fig. \ref{fig:interferometer_zoom} for zoom-ins of the boxed regions.
	Third column: FWHM = 7 arcsec ($42 h^{-1}$ kpc), representative of the resolution available with, e.g., \emph{Herschel} PACS
	and the current configuration of the LMT. Fourth column: FWHM = 15 arcsec ($89 h^{-1}$ kpc), representative of, e.g., \emph{Herschel} SPIRE and the JCMT.
	The fields of view are 23.5 arcsec ($140 h^{-1}$ kpc) for the first three columns and 59 arcsec ($351 h^{-1}$ kpc) for the last.
	For all but the first column, Gaussian noise with realistic amplitude has been added. The same
	noise map is used for all images in a given column. When present, scale bars show the beam FWHM, and contours correspond to $3\sigma$, $4\sigma$, $5\sigma$, $10\sigma$,
	and $20\sigma$. The first column has logarithmic scaling; all others are linear. Units are arbitrary.}
  \label{fig:merger_sequence}
\end{figure*}

\begin{figure}
  \centering
    \includegraphics[width=4cm]{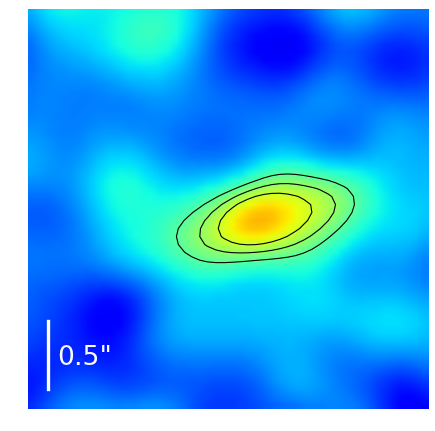}
    \includegraphics[width=4cm]{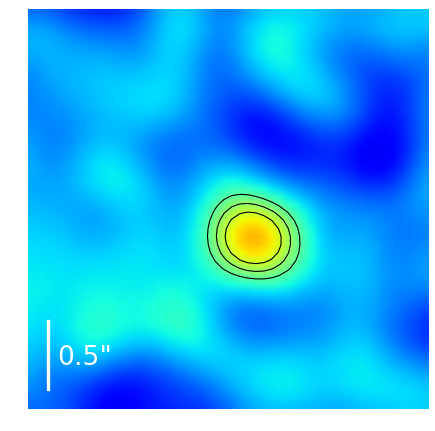} \\
    \includegraphics[width=4cm]{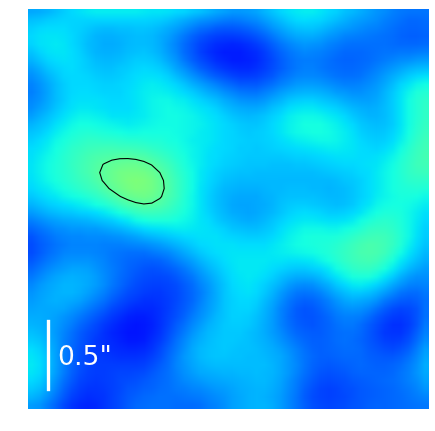} 
    \includegraphics[width=4cm]{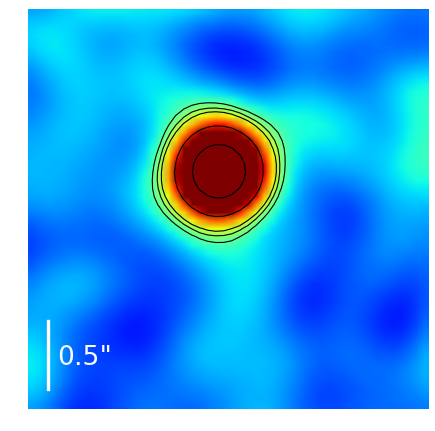} \\
    \includegraphics[width=5cm]{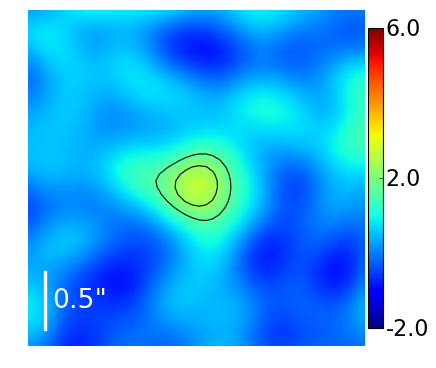}
  \caption{Zoom-ins of the regions marked with white boxes in the second column of Fig. \ref{fig:merger_sequence}. The upper left (right) panel shows the
  left (right) region from the first-row image of Fig. \ref{fig:merger_sequence}. The middle left corresponds to the second row, the middle right to the third row,
  and the bottom to the fourth row of Fig. \ref{fig:merger_sequence}. The FOV of each image is 3 arcsec ($18 h^{-1}$ kpc).
  See the caption of Fig. \ref{fig:merger_sequence} for further details.}
  \label{fig:interferometer_zoom}
\end{figure}

In \citet[][hereafter H11]{Hayward:2011smg_selection} we suggested a modification to the canonical picture, arguing that SMGs are not purely late-stage
major mergers but rather a heterogeneous
population, composed of: 1. late-stage major mergers undergoing strong starbursts; 2. early-stage major mergers (`galaxy-pair SMGs') which are quiescently
forming stars; 3. physically unrelated galaxies blended into one submm source \citep{Wang:2011}; and 4. isolated disc galaxies and minor mergers, which only
contribute significantly at the fainter end of the population.
One reason early-stage mergers also contribute is that observed submm flux increases rather weakly with SFR, and the starburst mode is significantly less efficient
at boosting submm flux than the quiescent mode. Physically, submm flux scales more weakly in starbursts for two reasons: 1. In high-redshift mergers significant star formation
occurs before the starburst induced at the coalescence of the two discs. During the starburst the `old' stars formed pre-burst 
can provide a non-negligible contribution to $\lir$. For example,
for the merger shown in the right panel of fig. 1 of H11, the stars formed pre-burst account for $\sim1/7$ of the total IR luminosity at the time of the burst.
Consequently, $\lir$ does not scale linearly with the instantaneous SFR during the burst because of the dust heating from the older stars (see, e.g., section
2.5 of \citealt{Kennicutt:1998review} and section 3.1 of H11).
2. Driven primarily by the strong drop in dust mass during the merger, the effective dust temperature of the SED increases sharply during the starburst.
This effect mitigates the increase in submm flux caused by the increased luminosity. These two effects result in a very weak scaling
($S_{850} \propto$ SFR$^{0.1}$, compared to $S_{850} \propto$ SFR$^{0.4}$ for the quiescently star-forming galaxies). For the
example in H11, an increase in SFR of $\sim 16\times$ causes a $\la 2\times$ boost in the submm flux.

The other reason early-stage mergers contribute significantly is that
the large (FWHM $\sim 15$ arcsec, or $\sim 120-130$ kpc at $z \sim 2-3$) beams of the single-dish submm telescopes used to detect SMGs
cause the two merging discs to be blended into a single source for much of the pre-coalescence stage of the merger. Fig. \ref{fig:merger_sequence}
demonstrates the effects of blending by showing observed-frame 850-$\micron$ (assuming $z = 2$) continuum images of the merger from H11 near apocentre
(first row), during final approach (second row), and at the peak of the starburst (third row). The fourth row shows the isolated disc from H11.
The first column shows the full-resolution simulated images. The others are meant to mimic the resolution attainable with various telescopes/interferometers. See
the figure caption for full details.

Comparison of the columns shows the importance of blending: For instruments with resolution $\sim 15$ arcsec (fourth column of Figure \ref{fig:merger_sequence}),
at all stages (even at apocentre) the two discs are blended into one submm source, and the morphologies of the galaxy-pair SMGs, the late-stage merger-induced
starbursts, and the isolated disc are similar. This resolution is typical of the surveys that have been done to date with SCUBA, AzTEC, and \emph{Herschel} SPIRE, so those
surveys should contain many galaxy-pair SMGs consisting of widely separated merging discs blended into one submm source.
With resolution $\sim 7$ arcsec (third column of Figure \ref{fig:merger_sequence}) the widely separated
galaxy-pair SMGs (first row) can be resolved, but some mergers which are nearer coalescence but still predominantly forming stars quiescently will
be blended into one source (second row). Thus wide-field surveys done with, e.g., \emph{Herschel} PACS or the Large Millimeter Telescope (LMT)
can more effectively distinguish galaxy-pair SMGs from starburst SMGs and should have a higher fraction of close pairs in the FIR/(sub)mm than
the SCUBA, AzTEC, and \emph{Herschel} SPIRE surveys. When the resolution is $\sim 0.5$ arcsec (second column of Figure \ref{fig:merger_sequence};
see Fig. \ref{fig:interferometer_zoom} for zoom-ins of the boxed regions),
typical of interferometers such as the SMA, the IRAM Plateau de Bure Interferometer, and ALMA, all but the mergers nearest coalescence are resolved
into two components, as this beam size corresponds to $\sim3 h^{-1}$ kpc at $z \sim 2-3$. At such low separations the mergers typically are undergoing strong
starbursts, so the SMGs observed as single components with interferometers should be predominantly starbursts or isolated discs. Note that when widely separated
merging discs are observed with some interferometers one of the sources may be outside the FOV, so from the interferometry such SMGs would appear to be isolated discs
whereas they are actually galaxy-pair SMGs. Clearly it is difficult to disentangle the SMG sub-populations using the number of components observed or the morphology.

The blending of the discs during the early stages of a merger is effective at creating SMGs because, by treating the system as a single source,
the integrated submm flux and SFR are both doubled; see sections 3.2 and 4.1 of H11 for further details. This blending, along with
the stronger scaling between submm flux and SFR in the quiescent mode compared to the starburst mode,
causes early-stage mergers to contribute significantly to the SMG population (details of the contribution will be presented in Hayward et al.,
in preparation). The brightest SMGs should still be merger-induced starbursts, but early-stage mergers must provide a significant contribution to the population.

There is already much evidence that some SMGs are early-stage mergers.
For example, \citet{Engel:2010} used submm interferometry to show that approximately half of their SMG sample are well-resolved binary systems. (See also
\citealt{Tacconi:2006,Tacconi:2008,Bothwell:2010,Riechers:2011b,Riechers:2011a}.)
Two of the twelve SMGs in the \citeauthor{Engel:2010} sample consist of two well-separated, resolved components (projected separations $\ga 20$ kpc),
and it is possible that they have missed one component of galaxy-pair SMGs with more widely separated components because of the limited field of view.
In such widely separated systems the star formation induced by the tidal torques exerted by the discs upon one another is not sufficient to drive a strong
starburst, so the discs would form stars at similar rates even if the companion were absent.
Thus such systems should be considered physically analogous to normal disc galaxies rather than late-stage mergers because they are forming stars via the quiescent
rather than starburst mode. In addition to the evidence from CO interferometry, support for the galaxy-pair contribution is provided by the frequency of multiple
counterparts in the radio \citep[e.g.,][]{Ivison:2002,Ivison:2007,Chapman:2005,Clements:2008,Younger:2009SMG_interf,Yun:2012}
and 24-\micron~ \citep[e.g.,][]{Pope:2006,Hwang:2010dust_T_evolution,Yun:2012} emission
and SMGs with morphologies that appear more like discs than late-stage mergers \citep[e.g.,][]{Bothwell:2010,Carilli:2010,Ricciardelli:2010,Targett:2011}.
(Note, however, that in gas-rich mergers discs can rapidly re-form, potentially confusing interpretation of these results; \citealt{Springel:2005disks,Narayanan:2008quasar_CO,
Robertson:2008,Hopkins:2009bulge_demo}.)

We have shown that physical arguments and observations suggest that the SMG population is a mix of quiescently star-forming galaxies and starbursts.
Thus SMGs differ significantly from local ULIRGs, which are exclusively starburst- or AGN-dominated, so one should draw comparisons
between the two populations with care. The heterogeneity complicates physical interpretation of the SMG population. For example, one should not apply
a CO--H$_2$ conversion factor appropriate for starbursts to the quiescently star-forming sub-population of SMGs.
Furthermore, proper treatment of all sub-populations is key for reproducing the observed SMG number counts
\citep[][Hayward et al., in preparation]{Hayward:2011num_cts_proc}.
However, the relative contributions of these various sub-populations is not yet observationally well-determined, and predictions of the relative contributions depend
sensitively on uncertain model details. We do not predict the relative contributions in the present work.
Instead, we wish to determine how one can observationally distinguish between starburst-driven (late-stage merger) SMGs and those powered by
quiescent star formation even when only integrated data are available. One can then use these diagnostics to observationally constrain the relative contributions
of starbursts and quiescently star-forming galaxies to the SMG population, thereby testing the bimodality we claim exists.

The rest of the paper is organised as follows:
We describe our simulation methodology in Section \ref{S:methods}. Section \ref{S:diagnostics} presents multiple observational
diagnostics that can be used to distinguish between
quiescently star-forming galaxies and starbursts from integrated data alone, including
the luminosity-effective dust temperature relation (Section \ref{S:L-T_relation}), star formation efficiency (Section \ref{S:SFE}),
IR excess (Section \ref{S:obscuration}), and the SFR--$\mstar$ relation (Section \ref{S:SFR-Mstar}).
In Section \ref{S:discussion} we discuss some implications of our work, and in Section \ref{S:conclusions} we summarise and conclude.

\section{Simulation methodology} \label{S:methods}

We analyse \gadgettwo \citep*{Springel:2001gadget,Springel:2005gadget}
3-D N-body/smoothed-particle hydrodynamics\footnote{Recent works \citep{Agertz:2007,Springel:2010arepo,Bauer:2011,Sijacki:2011}
suggest that the SPH method has several significant flaws, including artificially suppressing fluid instabilities, preventing efficient gas stripping of infalling structures,
and artificially damping turbulent eddies in the subsonic regime. Comparison of cosmological simulations run with \gadgettwo
to otherwise identical simulations run with the more accurate moving-mesh code \arepo \citep{Springel:2010arepo} shows that significant differences exist
even though all physics incorporated in the simulations is identical \citep{Vogelsberger:2011,Keres:2011,Torrey:2011}.
However, preliminary comparison of idealised merger simulations of the sort presented here
run with \gadgettwo and \arepo suggests that the two methods yield similar results (Hayward et al., in preparation).
Thus we expect our results to be robust in this regard.}
(SPH) simulations of mergers of equal-mass disc galaxies 
with the \sunrise \citep*{Jonsson:2006sunrise,Jonsson:2010sunrise} polychromatic Monte Carlo
dust RT code to calculate synthetic SEDs of the simulated galaxies.
The combination of \gadgettwo and \sunrise has been used to successfully reproduce
the SEDs/colours of a variety of galaxy populations, both low- and
high-redshift, including: local SINGS \citep{Kennicutt:2003,Dale:2007}
galaxies \citep{Jonsson:2010sunrise}; local ULIRGs \citep{Younger:2009}; extended UV discs \citep{Bush:2010};
24 \micron-selected galaxies \citep{Narayanan:2010dog}; massive, quiescent, compact $z \sim 2$ galaxies \citep{Wuyts:2009b,Wuyts:2010}; 
and K+A/post-starburst galaxies \citep{Snyder:2011}, among other populations. These successes support our application
of \gadget and \sunrise to modelling high-redshift ULIRGs.

\subsection{Hydrodynamic simulations}

{\sc Gadget-2}\footnote{A public version of \gadgettwo is available at \url{http://www.mpa-garching.mpg.de/~volker/gadget/index.html}.}
\citep{Springel:2001gadget,Springel:2005gadget} is a TreeSPH \citep{Hernquist:1989treesph} code that computes gravitational interactions via a
hierarchical tree method \citep{Barnes:1986} and gas dynamics via SPH \citep{Lucy:1977,Gingold:1977,Springel:2010}.
The formulation of SPH used explicitly conserves both energy and entropy when appropriate \citep{Springel:2002}.
Radiative heating and cooling is included in \gadgettwo following \citet*{Katz:1996}. Star formation is implemented using a volume-density-dependent KS
law \citep{Kennicutt:1998}, $\rho_{\rm SFR} \propto \rho_{\rm gas}^{1.5}$, with a minimum density threshold. The assumed KS index
results in a surface-density-dependent KS law consistent
with observations of $z \sim 2$ discs \citep{Krumholz:2007KS,Narayanan:2008CO_SFR,Narayanan:2011ks}, suggesting that it is reasonable
to use this prescription in our simulations of $z \sim 2$ mergers. The gas is enriched with metals assuming each particle behaves as a closed box,
so those gas particles with higher SFRs are more rapidly metal-enriched.

We use the sub-resolution two-phase ISM model of \citet{Springel:2003}. In this model, cold, dense clouds are embedded in a diffuse, hot medium.
Supernova feedback \citep{Cox:2006}, radiative heating and cooling, and star formation control the exchange of energy and mass in the two phases.
A simple model for black hole (BH) accretion and AGN feedback \citep*{Springel:2005feedback,DiMatteo:2005} is included.
BH sink particles with initial mass $10^5 \msun$ are included in both initial disc galaxies. They accrete via
Eddington-limited Bondi--Hoyle accretion \citep{Hoyle:1939,Bondi:1944}. The luminosity of each BH is calculated from the accretion rate
$\dot{M}_{\rm BH}$ assuming the radiative efficiency appropriate for a \citet{Shakura:1973} thin disc, 10 per cent. Thus
$\lbol = 0.1 \dot{M}c^2$. Five per cent of the luminosity emitted by the BHs is deposited into the surrounding ISM.

The simulations are initialised in the following manner: Exponential discs with initial gas fraction $f_g = \mgas/(\mstar+\mgas) = 0.8$\footnote{Note,
however, that we discard all snapshots with $f_g < 0.4$ for reasons described below.} are embedded in
dark matter haloes described by a \citet{Hernquist:1990} profile.
The progenitor discs are scaled to $ z \sim 3$ as described in \citet{Robertson:2006} so that the mergers
occur at $z \sim 2$. We have selected galaxy masses representative of the SMG population \citep[e.g.,][]{Michalowski:2011}.
The gravitational softening lengths are 200$h^{-1}$ pc for the dark matter particles and 100$h^{-1}$ pc for the stellar, gas,
and BH particles. Each disc galaxy is composed of $6 \times 10^4$ dark matter particles, $4 \times 10^4$ stellar particles,
$4 \times 10^4$ gas particles, and 1 BH particle. Two identical discs are initialised on parabolic orbits
with initial separation $R_{\rm init} = 5R_{\rm vir}/8$ and pericentric distance twice the disc scale length \citep{Robertson:2006}.
We analyse only equal-mass mergers because these simulations provide sufficient examples of quiescently star-forming galaxies
(during the early stages) and starbursts (near coalescence). The differences between quiescent and starburst modes
are insensitive to orbit and merger mass ratio. (Not all mergers induce strong starbursts, but those are irrelevant for our present purposes.)
The physical parameters of the simulated major mergers are summarised in Table \ref{tab:merger_properties}.

\ctable[
    caption	=	{Simulation parameters \label{tab:merger_properties}},
    			center,
			star
  ]{lcccccccccc}{
    \tnote[a]{Virial mass of each progenitor.}
    \tnote[b]{Initial stellar mass of each disc.}
    \tnote[c]{Initial gas mass of each disc.}
    \tnote[d]{Initial separation of the discs.}
    \tnote[e]{Pericentric passage distance.}
    \tnote[f]{Orientation of each disc's spin axis in spherical coordinates.}
    }{
																		\FL
      		& $M_{200}$\tmark[a]
		& $M_{\star, {\rm  init}}$\tmark[b]
		& $M_{\rm gas, init}$\tmark[c]
		& $R_{\rm init}$\tmark[d]
		& $R_{\rm peri}$\tmark[e]
		& $\theta_1$\tmark[f]
		& $\phi_1$\tmark[f]
		& $\theta_2$\tmark[f]
		& $\phi_2$\tmark[f]													\NN
      Name	& ($10^{12} h^{-1} \msun$)	& ($10^{10} h^{-1} \msun$)	& ($10^{10} h^{-1} \msun$)	& ($h^{-1}$ kpc)	&
      ($h^{-1}$ kpc)	& (deg)		& (deg)		& (deg)		& (deg)				\ML
	b6i & 6.2 & 5.3 & 22 & 70 & 6.7 & 0 & 0 & 71 & 30							\NN
	b6j & 6.2 & 5.3 & 22 & 70 & 6.7 & -109 & 90 & 71 & 90						\NN
	b6k & 6.2 & 5.3 & 22 & 70 & 6.7 & -109 & -30 & 71 & -30						\NN
	b6l & 6.2 & 5.3 & 22 & 70 & 6.7 & -109 & 30 & 180 & 0						\NN
	b6m & 6.2 & 5.3 & 22 & 70 & 6.7 & 0 & 0 & 71 & 90							\NN
	b6n & 6.2 & 5.3 & 22 & 70 & 6.7 & -109 & -30 & 71 & 30						\NN
	b6o & 6.2 & 5.3 & 22 & 70 & 6.7 & -109 & 30 & 71 & -30						\NN
	b6p & 6.2 & 5.3 & 22 & 70 & 6.7 & -109 & 90 & 180 & 0						\NN
	b5.5i & 3.2 & 2.7 & 11 & 57 & 5.3 & 0 & 0 & 71 & 30							\NN
	b5.5j & 3.2 & 2.7 & 11 & 57 & 5.3 & -109 & 90 & 71 & 90						\NN
	b5.5k & 3.2 & 2.7 & 11 & 57 & 5.3 & -109 & -30 & 71 & -30					\NN
	b5.5l & 3.2 & 2.7 & 11 & 57 & 5.3 & -109 & 30 & 180 & 0						\NN
	b5.5m & 3.2 & 2.7 & 11 & 57 & 5.3 & 0 & 0 & 71 & 90						\NN
	b5.5n & 3.2 & 2.7 & 11 & 57 & 5.3 & -109 & -30 & 71 & 30						\NN
	b5.5o & 3.2 & 2.7 & 11 & 57 & 5.3 & -109 & 30 & 71 & -30						\NN
	b5.5p & 3.2 & 2.7 & 11 & 57 & 5.3 & -109 & 90 & 180 & 0						\NN
	b5i & 1.6 & 1.4 & 5.6 & 44 & 4.0 & 0 & 0 & 71 & 30							\NN
	b5j & 1.6 & 1.4 & 5.6 & 44 & 4.0 & -109 & 90 & 71 & 90						\NN
	b5k & 1.6 & 1.4 & 5.6 & 44 & 4.0 & -109 & -30 & 71 & -30						\NN
	b5l & 1.6 & 1.4 & 5.6 & 44 & 4.0 & -109 & 30 & 180 & 0						\NN
	b5m & 1.6 & 1.4 & 5.6 & 44 & 4.0 & 0 & 0 & 71 & 90							\NN
	b5n & 1.6 & 1.4 & 5.6 & 44 & 4.0 & -109 & -30 & 71 & 30						\NN
	b5o & 1.6 & 1.4 & 5.6 & 44 & 4.0 & -109 & 30 & 71 & -30						\NN
	b5p & 1.6 & 1.4 & 5.6 & 44 & 4.0 & -109 & 90 & 180 & 0						\LL
    }

\subsection{Radiative transfer}

We use the 3-D Monte Carlo
dust RT code {\sc Sunrise}\footnote{\sunrise is publicly available at \url{http://code.google.com/p/sunrise/}.}
\citep{Jonsson:2006sunrise,Jonsson:2010sunrise} in post-processing
to calculate the far-UV--mm SEDs of each simulated merger at 10 Myr intervals.
We briefly describe the \sunrise calculation here, but the reader is encouraged
to see \citet{Jonsson:2010sunrise} for full details.
\sunrise uses the stellar and BH particles from the \gadgettwo simulations
as radiation sources. Each stellar particle with age $> 10$ Myr is treated as a single-age stellar population
and assigned a {\sc Starburst99} \citep{Leitherer:1999} SED template appropriate for its age and metallicity.
Stellar particles with age $< 10$ Myr are assigned a template from \citet{Groves:2008}, which includes emission
from the HII regions surrounding the clusters. We do not include the photo-dissociation regions for the reasons
discussed in detail in section 2.2.1 of H11.
The stars in the initial discs are assigned ages by assuming that the population was formed at a constant
rate equal to the SFR of the initial snapshot.
The gas and stars present in the initial discs have metallicity Z = 0.015,
which results in the galaxies being roughly on the $z \sim 2$ mass--metallicity relation during the starburst.
The BH particles are assigned SEDs using the luminosity-dependent
templates of \citet{Hopkins:2007}. These templates are derived from observations of un-reddened quasars,
so they include the intrinsic power-law emission and reprocessed hot dust emission from the torus.

\sunrise calculates the dust distribution by projecting the \gadgettwo gas-phase metal density
onto a 3-D adaptive-mesh-refinement grid using the SPH smoothing kernel and assuming a
dust-to-metal density ratio of 0.4 \citep{Dwek:1998,James:2002}. A minimum cell size of 55$h^{-1}$ pc is used;
this is sufficient to ensure the SEDs are converged to within $\la10$ per cent at
all wavelengths. Grain compositions, size distributions, and optical properties are given by the
Milky Way $R = 3.1$ dust model of \citet{Weingartner:2001} as updated by \citet{Draine:2007}. The opacity
curve for this model has a power-law slope in the far-IR $\beta \approx 2$.

To perform the RT we use $10^7$ photon packets for each stage, $\sim10 \times$ the number of grid cells.
This limits Monte Carlo noise to less than a few percent. \sunrise randomly
emits the photon packets from the sources and randomly draws interaction optical depths using the appropriate
probability distributions.
At the interaction optical depth a fraction of the photon packet's intensity is absorbed; the remainder is scattered
into a direction randomly drawn using the scattering phase function. This is repeated until the photon packet leaves the grid
or its intensity drops below a minimum value.

The energy absorbed by the dust is re-radiated in the IR. \sunrise assumes all dust
(except for half of the PAHs with grain size $<100$ \AA; see \citealt{Jonsson:2010sunrise} for details) is in thermal
equilibrium, so the dust temperature is calculated by setting the luminosity absorbed by each grain
equal to the energy emitted by the grain. The equilibrium temperature of a grain depends on the local
radiation field heating the grain and its absorption cross section, so there are in principle $n_{\rm cells} * n_{\rm grain~sizes}$
different dust temperatures in a given \sunrise calculation. This is important to keep in mind when one
considers fitting IR SEDs with modified blackbodies, as discussed below.

In high-density environments the ISM can be optically thick in the IR; this is common in the central
regions of the late-stage mergers modelled here. Consequently, one must account for attenuation of the dust
re-emission (aka dust self-absorption). Furthermore, since the IR emission absorbed heats the dust,
one must iterate the dust temperature calculation and RT of the dust emission until the dust temperatures are converged.
{\sc Sunrise} uses a reference field technique similar to that of \citet{Juvela:2005} to perform this iteration.
We encourage the interested reader to see \citet{Jonsson:2010sunrise} and \citet{Jonsson:2010gpu}
for details.

The \sunrise calculation yields spatially resolved, multi-wavelength (for these simulations there are 120 wavelengths
sampling the UV--mm range) SEDs for each galaxy snapshot observed from 7 different viewing angles distributed
uniformly in solid angle. The data are analogous to data yielded by integrated field unit (IFU) spectrographs.
For this work we spatially integrate to calculate integrated SEDs for the system. When calculating
observed flux densities we assume the simulated galaxies are at redshift $z = 2$.

\section{Observational diagnostics to distinguish between star formation modes} \label{S:diagnostics}

In this section we present multiple observational diagnostics that can distinguish between quiescently star-forming galaxies
(for SMGs this includes galaxy pairs and isolated discs) and starbursts induced at the final coalescence of major mergers.
We present diagnostics that rely only on integrated broadband photometry and CO line intensities (to determine gas mass).
We also assume that sufficiently accurate redshifts are known.
Spatially resolved data, such as that provided by (sub)mm interferometers and near-IR IFU spectrographs,
can potentially provide more diagnostic power but come at a much greater observational cost. However, even with high-resolution
near-IR data it can be difficult to distinguish between disc galaxies and gas-rich mergers in which discs re-form
shortly after final coalescence \citep{Robertson:2008},
and because of the high attenuation of SMGs near-IR observations may not probe the central starburst regions.
The FIR and (sub)mm are able to probe much deeper into the central regions, but the spatial resolution available at these
wavelengths is much coarser unless an interferometer is used, so typically only integrated FIR and (sub)mm photometry are available for high-redshift galaxies.
(However, ALMA will soon change this situation drastically.)
Finally, mergers and starbursts are not equivalent (see above), so identifying an object as a merger based on morphology does not ensure it
is also a starburst. Thus diagnostics that make use of only integrated data will continue to be crucial for distinguishing between star formation
modes and understanding the properties of high-redshift galaxies.

We have identified the starburst phase by defining the baseline SFR as the minimum SFR that occurs between first passage and
coalescence and selecting all snapshots where the instantaneous SFR is $>3\times$ that baseline SFR.
This factor is chosen so that the star formation induced by the merger dominates that which would occur in the discs
even if they were not merging. Increasing (decreasing) the threshold would result in less (more) sources identified as bursts
and amplify (diminish) the differences between modes that we describe below.
The snapshots that meet the starburst criterion are labelled `starburst' and plotted as blue squares.
Since the mutual gravitational torques are sub-dominant at first passage relative to internal
instabilities, the galaxies are primarily quiescently star-forming prior to the starburst induced at coalescence.
We thus label all snapshots before the starburst phase `quiescent' and plot them as black circles.
All snapshots after the starburst phase are neglected, as these are typically AGN or spheroids with relatively little ongoing star formation.
We will investigate the FIR properties of obscured AGN in detail in future work.

Our focus is the bright SMG population, defined by $S_{850} > 5$ mJy, so throughout this work simulated SMG data points are plotted with
larger symbols (black circles for quiescently star-forming and blue squares for starburst SMGs). However, in order to comment on the
applicability of the diagnostics to the high-redshift ULIRG population in general, data for simulated galaxies that would not be selected as SMGs
are also plotted. Small black circles (blue squares) correspond to quiescently star-forming (starburst) ULIRGs with $S_{850} < 5$ mJy.

We have neglected all snapshots with greater than 40 per cent gas fraction in order to eliminate the early parts of the simulations when the discs are
still stabilising and to remain consistent with (albeit uncertain; e.g., \citealt{Bothwell:2010}) observational constraints \citep{Tacconi:2006,Tacconi:2008}.
Such high initial gas fractions are required
to maintain sufficient gas until the time of coalescence because our simulations do not include any additional gas supply beyond what the galaxies start with.
We have checked that the results are qualitatively the same when we use an initial gas fraction of 60 per cent and include all snapshots.
All quantities plotted are totals for the entire system because we wish to present observational diagnostics based on integrated data alone.
In all figures, if a simulated galaxy sub-population has sample size $N > 100$ only a randomly selected subsample of 100 objects is plotted;
this is done to improve the readability of the plots, as a sub-sample of 100 is sufficient to show the distribution of the simulated data.
Finally, note that we plot data from idealised simulations without applying any weighting to account for cosmological abundances.
Thus the exact distribution of data in the various diagnostic plots is not necessarily representative of the SMG population.
What is meaningful, however, is when sub-populations occupy distinct regimes in a diagnostic plot;
this is a clear prediction for how the star formation modes should differ and how one can observationally disentangle the classes
in order to determine their relative contributions to a given galaxy population.

\subsection{Luminosity--effective dust temperature relation} \label{S:L-T_relation}

Far-IR (FIR) galaxy SEDs are often described in terms of a `dust temperature' obtained via fitting a single-temperature (single-$T$)
modified blackbody to the FIR SED \citep{Hildebrand:1983}. The single-$T$ modified blackbody form is
\begin{equation} \label{eq:mod_BB}
S_{\nu} = S_0 \left(1-e^{-(\lambda/\lambda_1)^{-\beta}}\right) B_{\nu}(\tdust),
\end{equation}
where $S_{\nu}$ is the flux density at rest-frame frequency $\nu$, $S_0$ is the normalisation, $\lambda_1$ is the wavelength at which
the effective optical depth $\tau_{\lambda} = 1$, $\beta$ is the effective slope of the emissivity in the FIR, $\tdust$ is the effective dust
temperature\footnote{Note, however, that in both the simulations and reality there is \textit{always} a distribution of dust temperatures,
as the temperature of a grain depends on the local radiation field and the size of the grain. The effective dust temperature
fit in this manner may provide a good approximation to the luminosity-weighted average dust temperature, but to our knowledge
this correspondence has not been demonstrated.}, and $B_{\nu}(\tdust)$ is the Planck function. The form is motivated by a simple
model: If it is assumed that the entire mass
of dust has \emph{physical} temperature equal to $\tdust$, the source has constant source function $B_{\nu}(\tdust)$. Assuming also that
the source is uniform, with projected area $A$ and angular size $d\Omega$, and that the opacity curve is a power law in the FIR,
\begin{equation} \label{eq:kappa}
\kappa_{\lambda} = \kappa_0 (\lambda/\lambda_0)^{-\beta},
\end{equation}
the optical depth is
\begin{equation} \label{eq:tau_lambda}
\tau_{\lambda} = \frac{\kappa_{\lambda} M_d}{A} = \frac{\kappa_{0} M_d}{d\Omega D^2}\left(\frac{\lambda}{\lambda_0}\right)^{-\beta},
\end{equation}
where $D$ is the distance to the source, $\mdust$ is the dust mass, and $\kappa_0$ is the opacity at wavelength
$\lambda_0$.\footnote{It is important to distinguish between $\lambda_0$ and $\lambda_1$. The former is just part of the parameterisation
of the opacity curve of the dust (see Equation \ref{eq:kappa}). The latter depends on the dust mass and effective area of the source in
addition to the dust opacity. $\lambda_1$ can be obtained from Equation (\ref{eq:tau_lambda}) by setting $\tau_{\lambda} = 1$ and solving for $\lambda$.}
(For simplicity we have assumed $z << 1$ here. In practice, we transform our simulation photometry to the rest frame before fitting.)
Under these assumptions, the solution of the radiative transfer equation is
\begin{eqnarray}
S_{\nu} &=& d\Omega \left(1-e^{-\tau_{\lambda}}\right)B_{\nu}(\tdust) \\
&=& d\Omega \left[1-\exp \left(\frac{-\kappa_0 M_d}{d\Omega D^2}\left(\frac{\lambda}{\lambda_0}\right)^{-\beta}\right)\right] B_{\nu}(\tdust). \label{eq:phys_mod_BB}
\end{eqnarray}
Note that the validity of Equation (\ref{eq:phys_mod_BB}) is limited to the IR wavelengths where the opacity curve has the assumed power-law form.
Typically it is assumed that optical depths in the FIR are small, so $\left[1-\exp \left({-(\lambda/\lambda_1)^{-\beta}}\right)\right] \approx (\lambda/\lambda_1)^{-\beta}.$ Thus
\begin{equation} \label{eq:mod_BB_ot}
S_{\nu} = S_0' \lambda^{-\beta} B_{\nu}(\tdust),
\end{equation}
where $S_0' = S_0 \lambda_1^{\beta}$. We shall refer to the form given in Equation (\ref{eq:mod_BB_ot}) as the
single-$T$ optically thin (OT) modified blackbody.
We refer the reader to the appendix of H11 for more details about the modified blackbody forms.

The simple model described above provides motivation for the fitting forms given in Equations (\ref{eq:mod_BB}) and (\ref{eq:mod_BB_ot}).
However, because the simple assumptions inherent in the model (specifically the single dust temperature and assumed geometry) are
certainly not true for real galaxies, one should not assume a priori that the parameters derived via fitting FIR SEDs with
Equations (\ref{eq:mod_BB}) or (\ref{eq:mod_BB_ot}) can be used to infer the physical quantities present in Equation (\ref{eq:phys_mod_BB}).
Indeed, the parameters determined using this fitting method should not be interpreted physically for multiple reasons: For example,
$\beta$ and $\tdust$ are degenerate \citep[e.g.,][]{Sajina:2006}, and
the $\tdust$ and $\beta$ one derives from the fitting depends strongly on observational noise \citep{Shetty:2009a}, dust temperature variations
along the line-of-sight \citep{Shetty:2009b}, and the wavelength range spanned by the photometry included in the fit (\citealt{Magnelli:2012}, hereafter M12).
Even in the case of a uniform radiation field grains of different sizes will have different temperatures, so there is
no reason to expect $\beta$ derived from the single-$T$ modified blackbody to be equal to the intrinsic
power-law index of the dust emissivity in the FIR. (However, if a distribution of dust temperatures is permitted it is
perhaps possible to recover the true $\beta$ from fitting the SED.) Furthermore, adding a significant component of very cold
dust can mimic the effect of high effective optical depth, so $\tau$ and the temperature distribution are also
degenerate \citep{Papadopoulos:2010}.
Finally, even the fitting method can cause incorrect conclusions: when traditional non-hierarchical $\chi^2$ minimisation is used,
observational noise can introduce a spurious anti-correlation between $\tdust$ and $\beta$ even when the true values of
$\tdust$ and $\beta$ are positively correlated; it is possible to avoid this pitfall by utilising a hierarchical Bayesian method \citep{Kelly:2012}.
Thus, at the least, one should use caution when attempting to infer physical conditions from the parameters derived by modified
blackbody fitting, as we demonstrate below.

However, even if the model parameters cannot be interpreted physically, the modified blackbody forms often provide acceptable
descriptions of the FIR SEDs of galaxies. Thus in order to compare to observations in a meaningful way and provide testable predictions from our models
we have fit the FIR SEDs of our simulated galaxies with both the optically thin and full forms of the modified blackbody. We have also used a more
sophisticated model which assumes a power-law distribution of temperatures. The results for each fitting form are discussed in turn below.

\subsubsection{Single-$T$, OT modified blackbody}

\begin{figure}
\centering
\plottwo{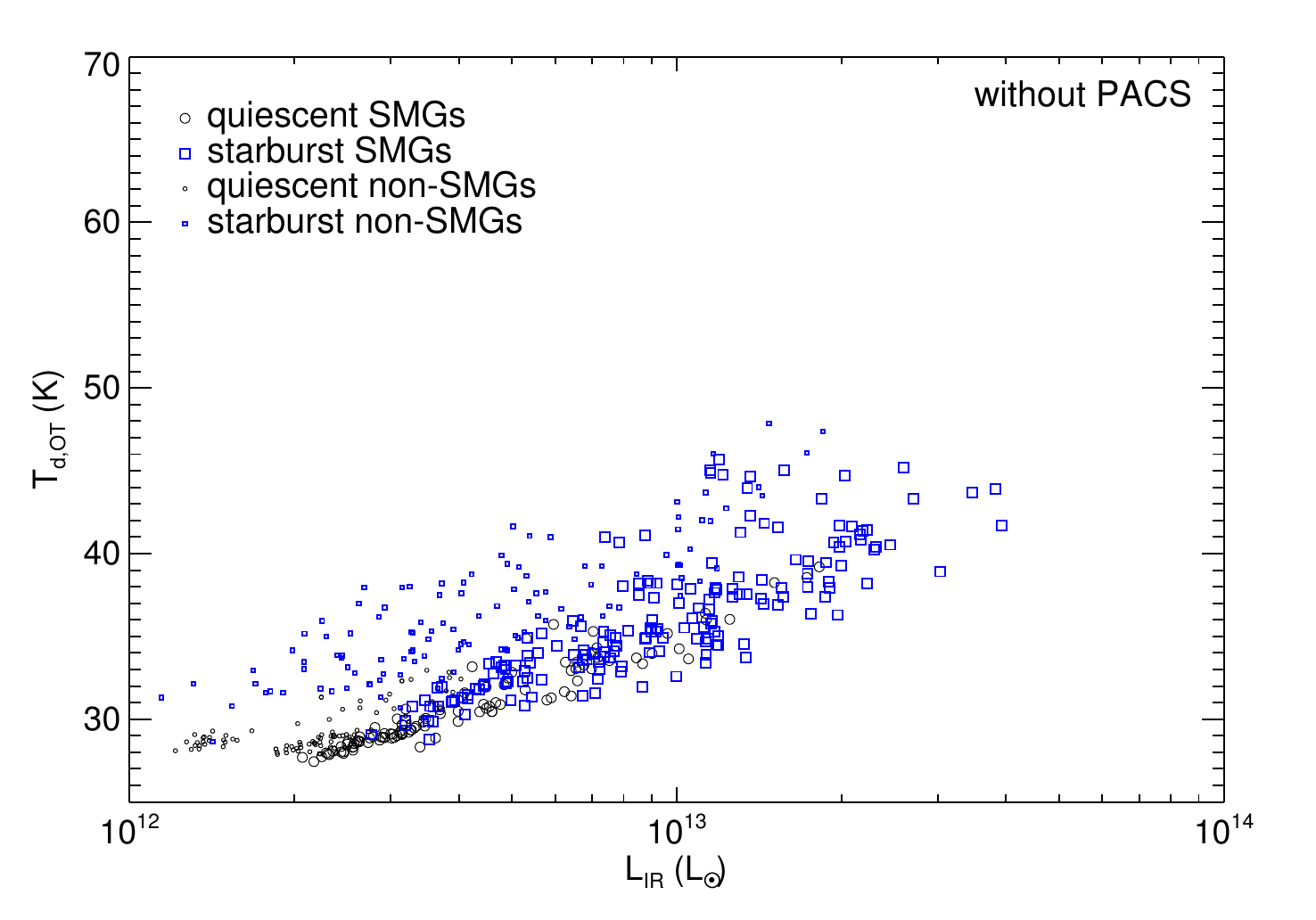}{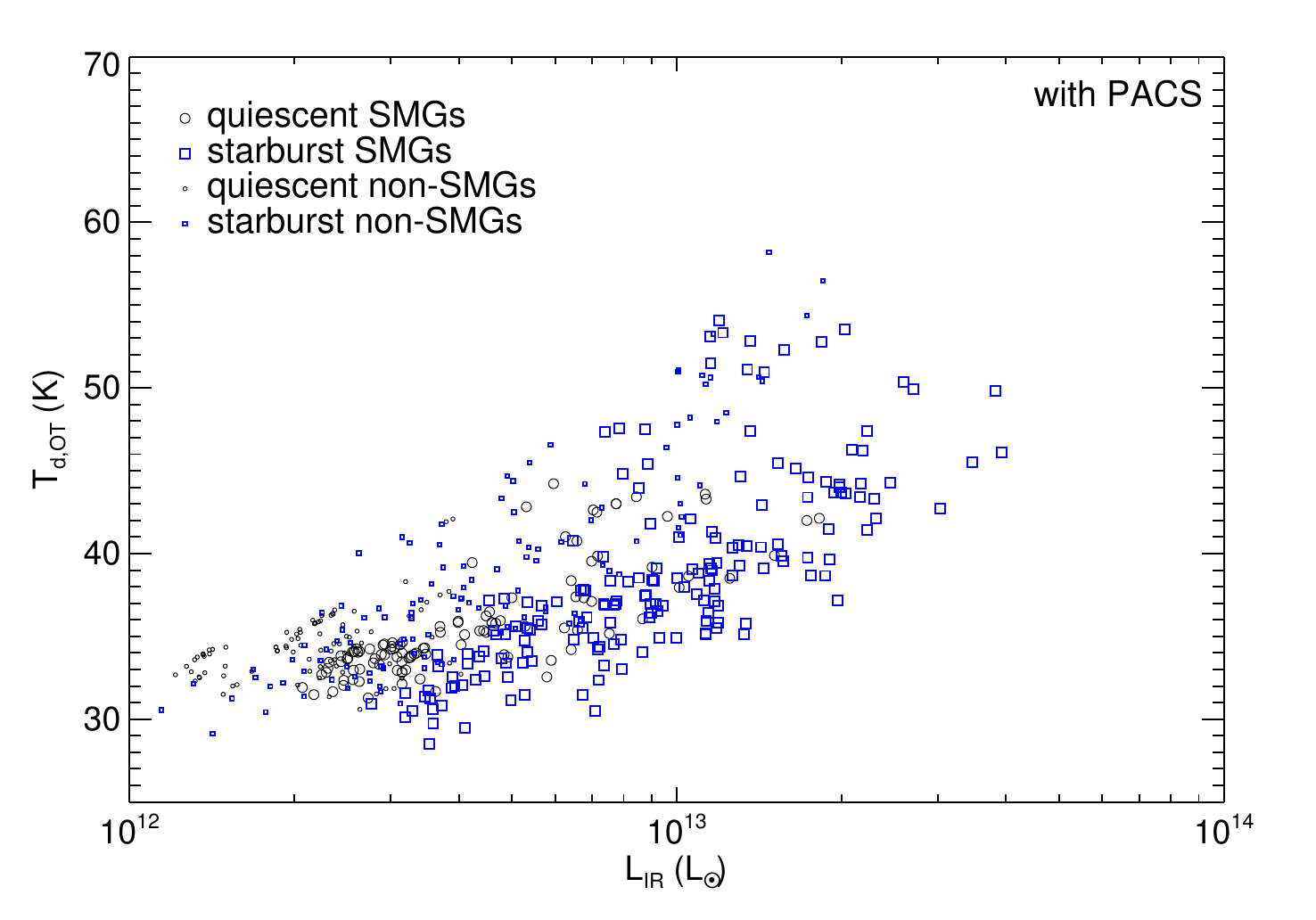}
\caption{Effective dust temperature $\tdust$ (K) derived from fitting the single-$T$ OT modified blackbody (Equation \ref{eq:mod_BB_ot}) to the simulated
SPIRE, SCUBA, and AzTEC photometry (top) and PACS 160-\micron, SPIRE, SCUBA, and AzTEC photometry (bottom) versus total IR luminosity $\lir$ ($\lsun$).
For the SMG population, effective dust temperature correlates strongly with luminosity, and the sources in the high-$\tdust$, high-$\lir$ region of the plot are almost exclusively
merger-induced starbursts. The SMG selection is biased against galaxies with hotter SEDs, so, for a given luminosity, the non-SMGs have higher $\tdust$ than the SMGs.
At the lower luminosities there is a significant population of galaxies missed by the SMG selection; these cause the $\tdust-\lir$ relation for the full population
to have larger scatter than that for the SMG population. Note also that adding the PACS 160-\micron ~data results in higher $\tdust$ and increased scatter.}
\label{fig:L-T_ot}
\end{figure}

\ctable[
    caption	=	{Median $\tdust$ for the OT single-$T$ modified blackbody \label{tab:Td_OT}},
    			center,
			star,
			doinside=\small
  ]{lcccccc}{
  }{
																																			\FL
      					& \multicolumn{2}{c}{without PACS 160-\micron} 	& \multicolumn{2}{c}{with PACS 160-\micron} 		& \multicolumn{2}{c}{sample size}				\NN
      Galaxy type			& $\lir < 10^{13} \lsun$	& $\lir \ge 10^{13} \lsun$	& $\lir < 10^{13} \lsun$	& $\lir \ge 10^{13} \lsun$ 	& $\lir < 10^{13} \lsun$	& $\lir \ge 10^{13} \lsun$ 	\ML
      Quiescent SMGs		&	30.1 K 			& 36.4				& 34.2				& 42.0				& 954				& 8					\NN
      Starburst SMGs		&	33.4				& 38.1				& 35.1				& 41.0				& 409				& 292				\NN
      Quiescent non-SMGs	&	28.9				& --					& 34.1				& --					& 648				& --					\NN
      Starburst non-SMGs	&	35.2				& 42.0				& 36.4				& 50.2				& 741				& 21					\LL
}

Fig. \ref{fig:L-T_ot} shows the distribution of our simulations on the $\tdust-L_{\rm IR}$ plot when the single-$T$ OT modified blackbody form (Equation \ref{eq:mod_BB_ot})
is used. The top panel shows $\tdust$ derived from fitting the simulated \textit{Herschel} SPIRE \citep{Griffin:2010} 250-, 350-, and 500-$\micron$,
SCUBA \citep{Holland:1999} 850-$\micron$, and AzTEC \citep{Wilson:2008} 1.1-mm photometry; on the bottom the
simulated \textit{Herschel} PACS \citep{Poglitsch:2010} 160-$\micron$ photometry are also included in the fit. We have excluded the PACS 70- and 100-\micron ~points
because often an acceptable fit cannot be obtained if those points are used. Possible reasons for this are that, at those wavelengths,
stochastically heated grains contribute significantly to the SED and the assumption of optical thinness is violated most.
When performing the Levenberg--Marquardt least-squares fit
we have assumed 10 K $\le \tdust \le$ 100 K, $1.0 \le \beta \le 2.5$, and constant fractional flux error of ten per cent.
The median effective dust temperatures for the each galaxy type are given in Table \ref{tab:Td_OT}, where the galaxies have been divided into
two bins, $\lir < 10^{13} \lsun$ and $\lir \ge 10^{13} \lsun$.

The key trends to take away from Fig. \ref{fig:L-T_ot} are that effective dust temperature correlates with luminosity, and the most luminous,
hottest SMGs are almost exclusively starbursts. When the PACS photometry are (not) used, almost all SMGs with
$\tdust \ga$ 40 (35) K are starbursts. Thus one can use a cut in $\tdust$ to cleanly select starbursts from the overall SMG population. Furthermore,
for the SMG population the cut $\lir > 10^{13} \lsun$ can also be used to cleanly select starburst SMGs.

When the non-SMG population is also considered the scatter in the $\tdust-\lir$ relation is increased, as there is a significant population of galaxies
at the lower-$\lir$ end with relatively high $\tdust$ that are missed by the SMG selection (see also fig. 1 of M12).
These simulated galaxies correspond to the observed `hot-dust ULIRGs'
(\citealt{Chapman:2004,Chapman:2008,Casey:2009,Casey:2010,Magnelli:2010}; M12); we discuss the differences between the SMG and non-SMG sub-populations in detail below
(Section \ref{S:smg_vs_non}).
At a given $\lir$ the non-SMGs have higher effective dust temperatures than the SMGs because of the bias inherent in the SMG selection.\footnote{Note that this is not
necessarily reflected in the tables of median temperatures because the non-SMGs in a given bin tend to have lower $\lir$ than the SMGs; thus the comparison of the
median $\tdust$ values for the bins is not the same as comparing the values for populations at fixed $\lir$.}
As for the SMG population, almost all non-SMGs with $\tdust \ga$ 40 (35) K when the PACS 160-\micron ~point is (not) used in the fit are starbursts.
Thus the $\tdust$ cut is an effective means for selecting starbursts from the overall galaxy population even at luminosities where there are a mix of starbursts
and quiescently star-forming galaxies. This is consistent with the observations of \citet{Magnelli:2010} and M12.

Note also that inclusion of the PACS photometry results in both increased dust temperature and larger scatter. This effect occurs because
we allow $\beta$ to vary; if $\beta$ is fixed then $\tdust$ is relatively similar whether or not the PACS photometry is included (\citealt{Magdis:2010dust_T}; M12).
The fact that the results of IR SED fitting can depend strongly on the regions of the SED sampled is another reason one must use caution when interpreting the
derived parameters. This is especially significant when comparing low- and high-redshift samples, as in this case a given observed data point can correspond to
significantly different rest-frame wavelengths for different redshifts.

\begin{figure}
\plotone{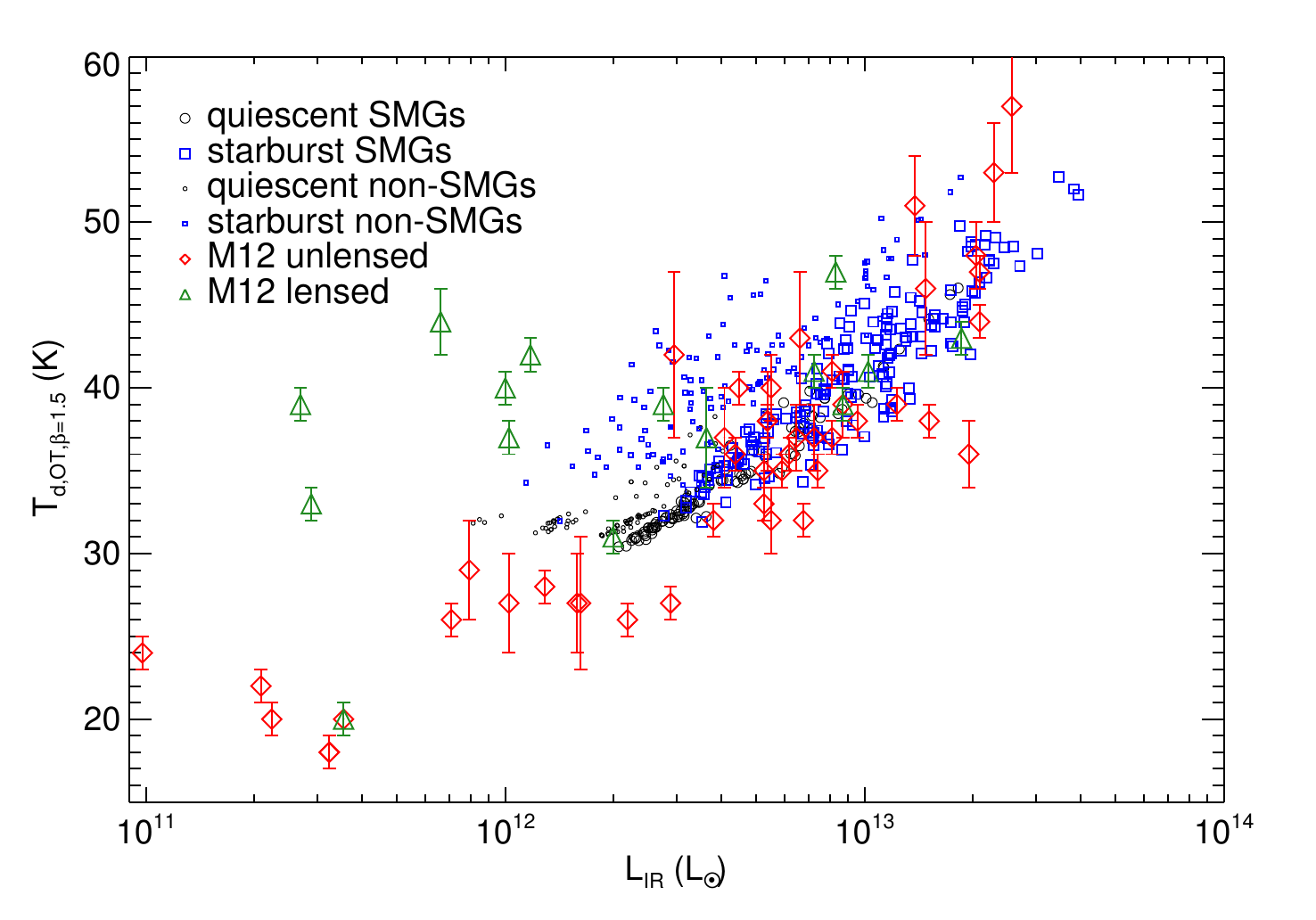}
\caption{Effective dust temperature (K) yielded by fitting the single-$T$ OT form of the modified blackbody with $\beta = 1.5$ fixed versus $\lir$ ($\lsun$).
In order to compare directly to M12 only the PACS 160-\micron, SPIRE, SCUBA, and AzTEC photometry have been used in the fit.
The large black circles (blue squares) are the simulated quiescently star-forming (starburst) SMGs, and the small black circles (blue squares) are the
simulated quiescently star-forming (starburst) non-SMGs. The red diamonds (green triangles) with error bars are the unlensed (lensed) SMGs ($S_{850} > 3$ mJy)
from M12. The agreement between the simulated and observed unlensed SMGs is very good for the entire luminosity range sampled by our simulations,
giving confidence that the FIR SEDs of the simulated galaxies are reasonable.}
\label{fig:L-T_obs}
\end{figure}

In Fig. \ref{fig:L-T_obs} we compare our simulations to the observations of M12. The effective temperature plotted here has
been determined by fitting the single-$T$ OT form with fixed $\beta = 1.5$, which is what \citeauthor{Magnelli:2012} have done. Only the PACS 160-$\micron$, SPIRE, SCUBA, and AzTEC
photometry have been used for the fits. For the range in $\lir$ covered by the simulations the agreement between the simulated and observed SMGs is very good.
This agreement gives confidence that the FIR SEDs of the simulated galaxies are reasonable. However, there are a few galaxies that lie below the region spanned
by the simulations. This is consistent with the conclusion of \citet{Jonsson:2010sunrise} that real galaxies form a more diverse family of SEDs than are generated by the simulations.
 
\subsubsection{Full form of the single-$T$ modified blackbody}

\begin{figure}
\centering
\plottwo{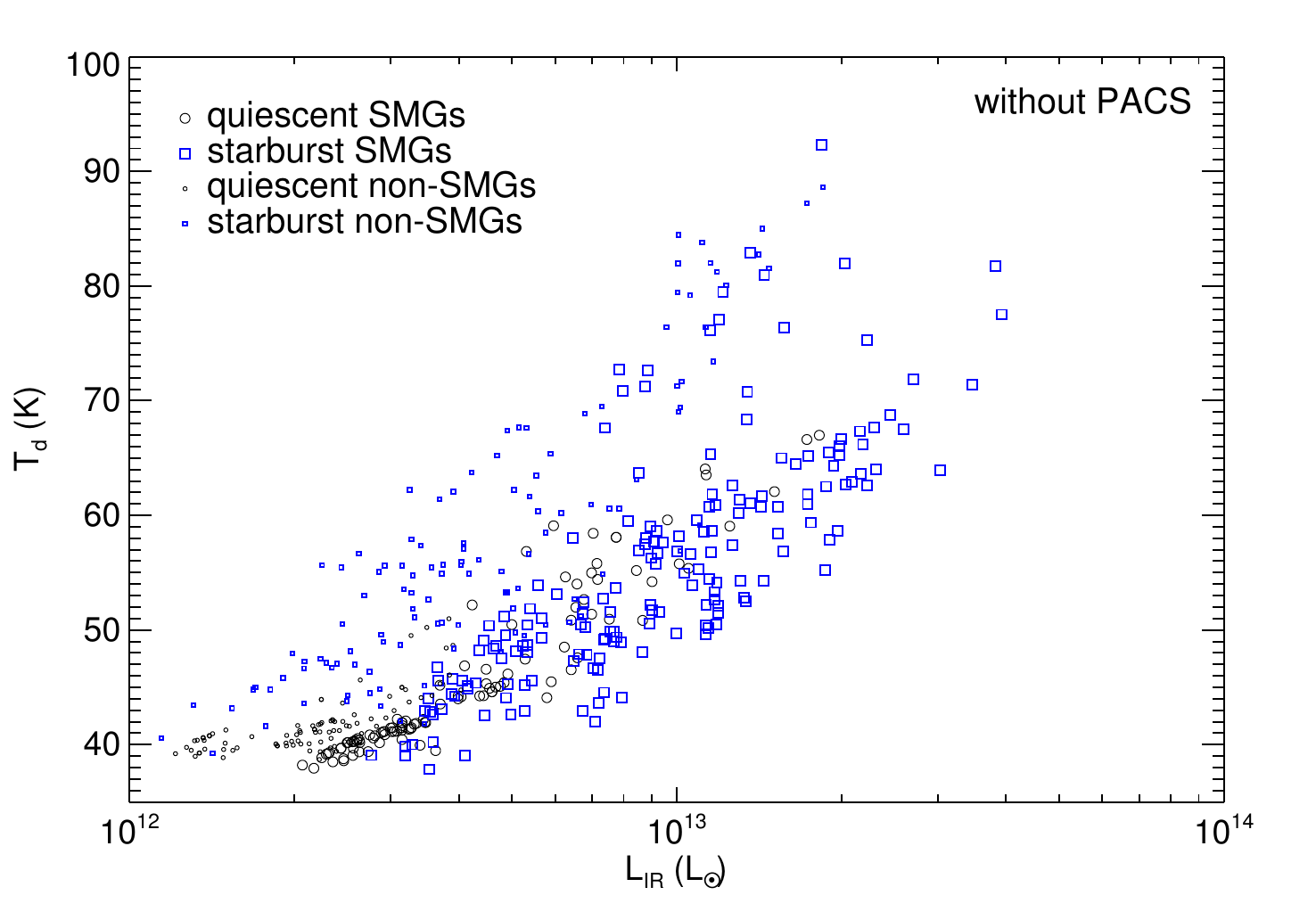}{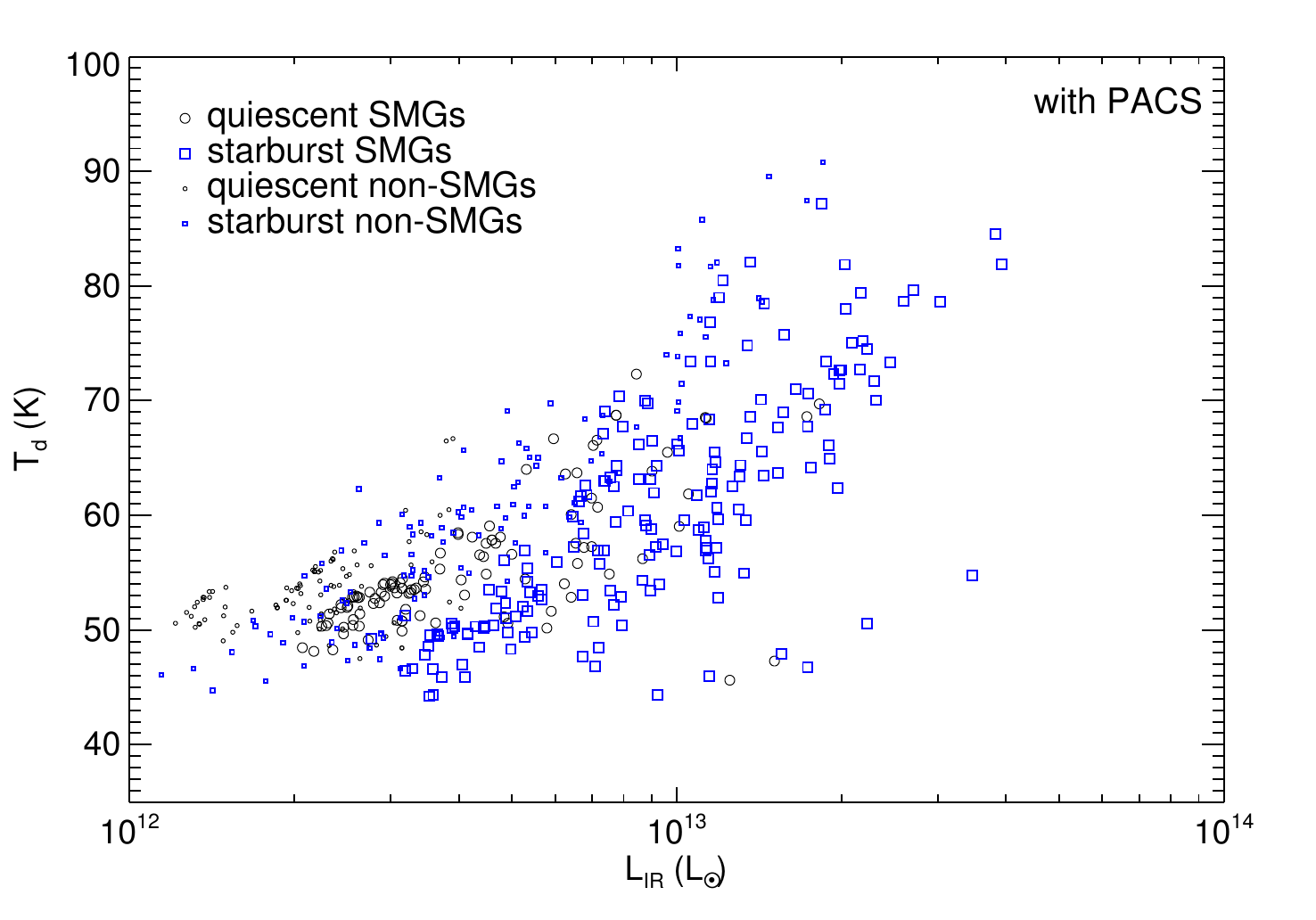}
\caption{Same as Fig. \ref{fig:L-T_ot}, but the effective dust temperature has been derived by fitting the full form of the modified blackbody (Equation \ref{eq:mod_BB}).
The $\tdust$ values derived using this form are systematically higher than when optical thinness is assumed (see text for details), but all trends -- except for the increased
scatter when PACS 100- and 160-\micron~data are included -- seen in Fig. \ref{fig:L-T_ot} still hold.}
\label{fig:L-T}
\end{figure}

\ctable[
    caption	=	{Median $\tdust$ for the full form of the single-$T$ modified blackbody \label{tab:Td_full}},
    			center,
			star,
			doinside = \small
  ]{lcccccc}{
  }{
																																					\FL
      					& \multicolumn{2}{c}{without PACS 100- \& 160-\micron} 	& \multicolumn{2}{c}{with PACS 100- \& 160-\micron} 	& \multicolumn{2}{c}{sample size}				\NN
      Galaxy type			& $\lir < 10^{13} \lsun$	& $\lir \ge 10^{13} \lsun$		& $\lir < 10^{13} \lsun$	& $\lir \ge 10^{13} \lsun$ 		& $\lir < 10^{13} \lsun$	& $\lir \ge 10^{13} \lsun$ 	\ML
      Quiescent SMGs		&	42.7	K			& 63.5					& 60.4				& 69.6					& 954				& 8					\NN
      Starburst SMGs		&	48.6				& 61.3					& 57.2				& 70.4					& 409				& 292				\NN
      Quiescent non-SMGs	&	40.9				& --						& 60.1				& --						& 648				& --					\NN
      Starburst non-SMGs	&	53.3				& 80.1					& 60.0				& 77.8					& 741				& 21					\LL
}

There is increasing evidence that the simple single-$T$ OT modified blackbody form given in Equation (\ref{eq:mod_BB_ot}) provides a poor fit
to the FIR SEDs of simulated (H11) and observed (\citealt{Lupu:2010}; \citealt{Conley:2011}; M12; Sajina et al., submitted)
high-redshift ULIRGs on the Wien side of the SED. Instead, more sophisticated forms, such as
Equation (\ref{eq:mod_BB}) or multiple-temperature models (e.g., \citealt{Dale:2002}; \citealt*{Clements:2010}; \citealt{Kovacs:2010}), 
must be used. Equation (\ref{eq:mod_BB}) allows non-negligible optical depths in the FIR, and allowing $\beta$ to vary mimics the effect of a temperature distribution
\citep[e.g.,][]{Shetty:2009b,Clements:2010}. The multiple-temperature models can account for non-negligible optical depths in the IR and multiple temperatures of dust,
both of which are physically more valid assumptions than those implicit in the single-$T$ OT blackbody model, but require at least one parameter more than Equation (\ref{eq:mod_BB}).

Fig. \ref{fig:L-T} shows the $\tdust-L_{\rm IR}$ plot when we derive the effective dust temperature $\tdust$ by fitting the full form of the modified blackbody,
Equation (\ref{eq:mod_BB}), to the simulated SPIRE+SCUBA+AzTEC (top) and PACS+SPIRE+SCUBA+AzTEC (bottom) photometry.
The PACS 70-$\micron$ data have not been used because at $z = 2$ observed-frame 70 \micron ~corresponds to rest-frame 23 \micron, a region of the SED dominated by
emission from stochastically heated grains. If the 70-\micron ~data are included an acceptable fit is often not possible.
We have assumed 10 K $\le \tdust \le$ 100 K, $1.0 \le \beta \le 2.5$, and ten per cent flux uncertainty.
$\lambda_1$ is allowed to vary freely, but in practice it is always greater than $\sim 375 ~\micron$. The median effective dust temperatures are given in Table \ref{tab:Td_full}.

Almost all trends demonstrated by Fig. \ref{fig:L-T_ot} hold here: effective dust temperature correlates with luminosity, and the most luminous,
hottest sources ($\tdust \ga$ 70 (60) K when the PACS data are (not) used) are almost exclusively starbursts. Almost all galaxies with
$\lir > 10^{13} \lsun$ are starbursts, but, again, a $\tdust$ cut is significantly better than an $\lir$ cut at selecting starbursts from the non-SMG population.
There is a significant population of galaxies at the lower luminosity end that are missed by the SMG selection because of their relatively high effective dust temperatures.
Finally, inclusion of the PACS photometry again results in higher effective dust temperature.

Comparison of Figs. \ref{fig:L-T_ot} and \ref{fig:L-T} and the median values given above shows that assuming optical thinness results in systematically lower $\tdust$ than
when Equation (\ref{eq:mod_BB}) is used. This occurs because $\left[1-\exp \left(-(\lambda/\lambda_1)^{-\beta}\right)\right] < (\lambda/\lambda_1)^{-\beta}$ for all $\lambda > 0$.
Thus, for fixed $\tdust$ and $\beta$, the assumption  of optical thinness will systematically over-predict the flux at frequencies for which $\tau \ga 1$. As a result, for a given SED
and fixed $\beta$, $\tdust$ derived from Equation (\ref{eq:mod_BB_ot})
will be lower than that derived from Equation (\ref{eq:mod_BB}). This effect has been demonstrated when fitting SEDs of high-redshift ULIRGs
(\citealt{Lupu:2010, Conley:2011}; Sajina et al., submitted), and it shows that one should use caution when attempting to interpret $\tdust$ physically.

\subsubsection{Power-law $T$-distribution model}

\begin{figure}
\centering
\plotone{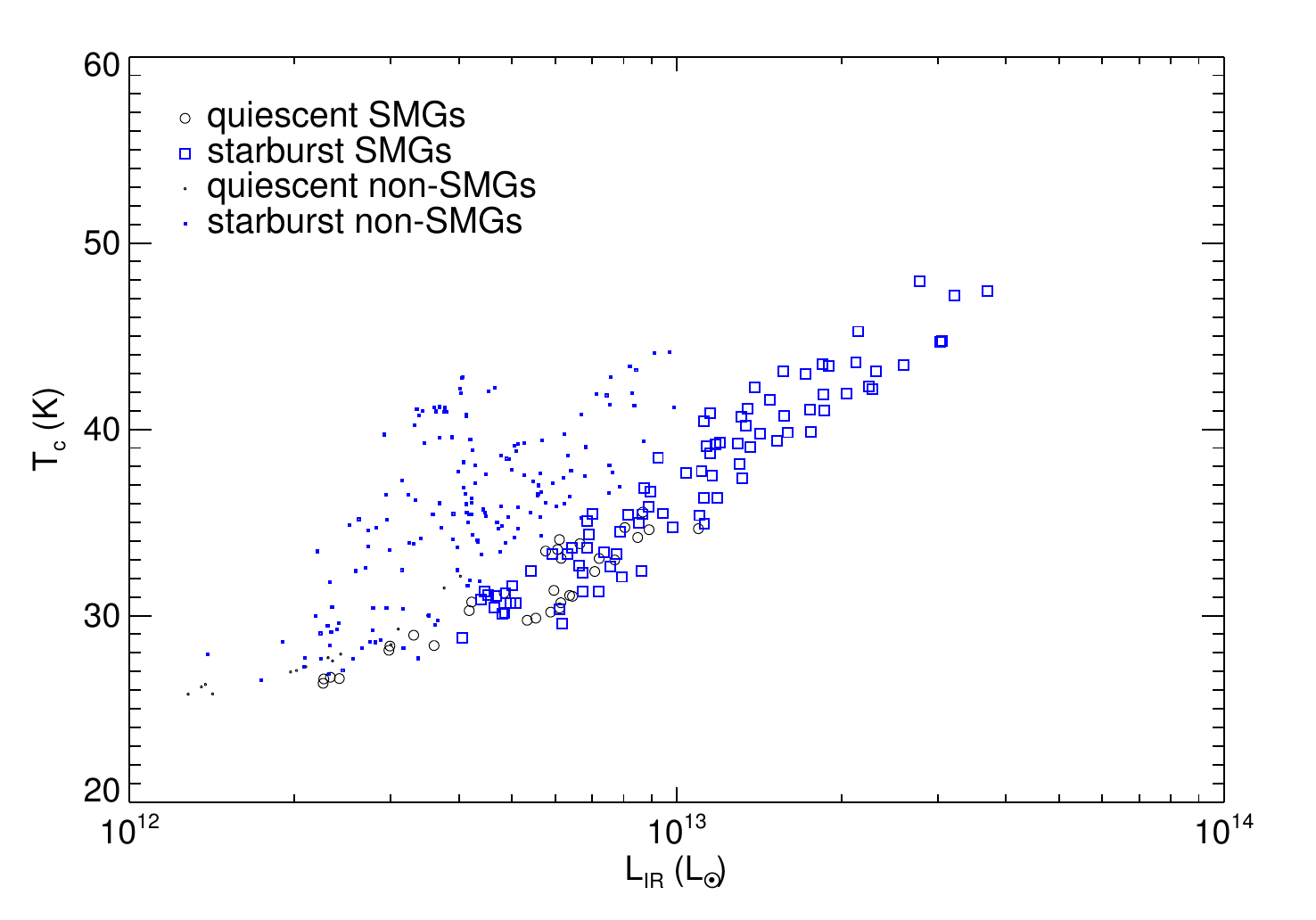}
\caption{Same as Fig. \ref{fig:L-T_ot}, but the effective dust temperature has been derived by assuming a power-law distribution of dust temperatures following
\citeauthor{Kovacs:2010} (\citeyear{Kovacs:2010}; the temperature plotted is the low-temperature cutoff of the power-law distribution -- see text for details).
Again, the dust temperatures differ,
but the trends are insensitive to the manner in which the effective dust temperature is derived. All sources with $\tc > 36$ K are starbursts.}
\label{fig:L-T_pl}
\end{figure}

\ctable[
    caption	=	{Median low-temperature cut-off $T_c$ for the power-law dust temperature distribution model \label{tab:Td_PL}},
    			center,
			star
  ]{lcccc}{
    \tnote[a]{The sample sizes here differ from those above because only a subset of the simulated galaxy SEDs have been fit with the
      power-law $T$-distribution form.}
  }{
      																										\FL
					& \multicolumn{2}{c}{with PACS 100- \& 160-\micron} 	& \multicolumn{2}{c}{sample size\tmark[a]}		\NN
      Galaxy type			& $\lir < 10^{13} \lsun$	& $\lir \ge 10^{13} \lsun$		& $\lir < 10^{13} \lsun$	& $\lir \ge 10^{13} \lsun$	\ML
      Quiescent SMGs		&	31.0	K			& 34.7					& 31					& 1					\NN
      Starburst SMGs		&	32.6				& 40.9					& 41					& 46					\NN
      Quiescent non-SMGs	&	27.6				& --						& 15					& --					\NN
      Starburst non-SMGs	&	36.0				& --						& 164				& --					\LL
}

We have also fit a subset
of the simulated galaxies' SEDs assuming a power-law temperature distribution with a low-temperature cutoff
following \citet{Kovacs:2010}. The fitting method is summarised here, but the reader is referred to \citet{Kovacs:2010} and
M12 for full details of the model. In this model, the dust has a distribution of physical temperatures given by $d\mdust/dT \propto T^{-\gamma}$ for $T > \tc$
and $\mdust(T) = 0$ otherwise. The effective optical depth $\tau_{\lambda}$ is parameterized as
\begin{equation}
\tau_{\lambda} = \kappa_0 \left(\frac{\lambda}{\lambda_0}\right)^{-\beta} \frac{\mdust}{\pi R^2_{\rm eff}},
\end{equation}
where $R_{\rm eff}$ can be thought of as an effective radius of the source.
Because of the added parameter $\gamma$ we have always used the PACS 100- and 160-$\micron$, SPIRE, SCUBA, and AzTEC data.
Following M12, we have assumed that single values of $\beta$, $\gamma$, and $R_{\rm eff}$ can be used for all sources, and
we have used a subset of 20 simulated galaxies to fix those parameters in the following manner: We first gridded the
$(\beta, \gamma, R_{\rm eff})$ parameter space. For each point in the grid we fit all 20 sources allowing $\tdust$ and $\mdust$
to vary. We summed the $\chi^2$ values of the individual fits for each parameter combination and chose the parameter combination with the
lowest total $\chi^2$ value. As above, the fractional flux error assumed is ten per cent. The parameters we determined in this manner are $(\beta, R_{\rm eff}, \gamma) =$
(1.6, 2 kpc, 8.7). \citet{Kovacs:2010} found that the parameters $(\beta, R_{\rm eff}, \gamma) =$ (1.5, 1 kpc, 6.7) provided the best description of their 
sample of $z \sim 2$ starbursts. For the SMG sample of M12 the best-fitting parameters are
$(\beta, R_{\rm eff}, \gamma) =$ ($2.0 \pm 0.2$, $3 \pm 1$ kpc, $7.3 \pm 0.3$).
The values of $\beta$ and $R_{\rm eff}$ derived from our simulations lie between those from the two observational studies.
However, our temperature distribution is steeper than those found by both \citeauthor{Kovacs:2010} and \citeauthor{Magnelli:2012},
perhaps because our simulations do not yet include stochastically heated very small grains and thus may underestimate the amount of dust at high temperatures.
We defer a detailed comparison of simulated and observed ULIRG SEDs and the derived dust parameters to future work.

The resulting $\tc-\lir$ plot is shown in Fig. \ref{fig:L-T_pl}.
Note that the temperature plotted here is the low-temperature cutoff, which is also the temperature of most of the dust because of the steepness of the power-law distribution.
The median $\tc$ values are given in Table \ref{tab:Td_PL}.
As seen for both single-$T$ fitting forms, for SMGs there is a clear correlation between effective dust temperature and luminosity, and the starbursts are the most luminous
and have the highest values of $\tc$. An effective dust temperature cut is very effective here: all galaxies with $\tc > 36$ K are starbursts.
Furthermore, at the lower-luminosity end there is again a population of starbursts missed by the SMG selection because of their relatively high effective dust temperatures.
Consequently, a $\tc$ cut can select a larger subset of the starbursts than can an $\lir$ cut.

M12 argue that the cut $\tc \ga 25$ K can very effectively separate starbursts (identified observationally by their offset from the SFR--$\mstar$ relation)
from quiescently star-forming SMGs, as we also see in our simulations. However, the value of $\tc$ that separates simulated starburst SMGs from quiescently star-forming SMGs
differs significantly from that derived by M12. The primary reason for this
discrepancy is that the $(\beta, R_{\rm eff}, \gamma)$ values derived for our simulations differ from those for the M12 sample. If the simulated
SEDs are fit with the $(\beta, R_{\rm eff}, \gamma)$ values from M12, the resulting $\tc$ values are significantly lower, the SED fits are still acceptable,
and the agreement between the observations and simulations is much improved. See section 6 of M12 for further details.

\subsubsection{Summary of results that are independent of the fitting form}

\begin{figure}
\centering
\includegraphics[width={0.90\columnwidth}]{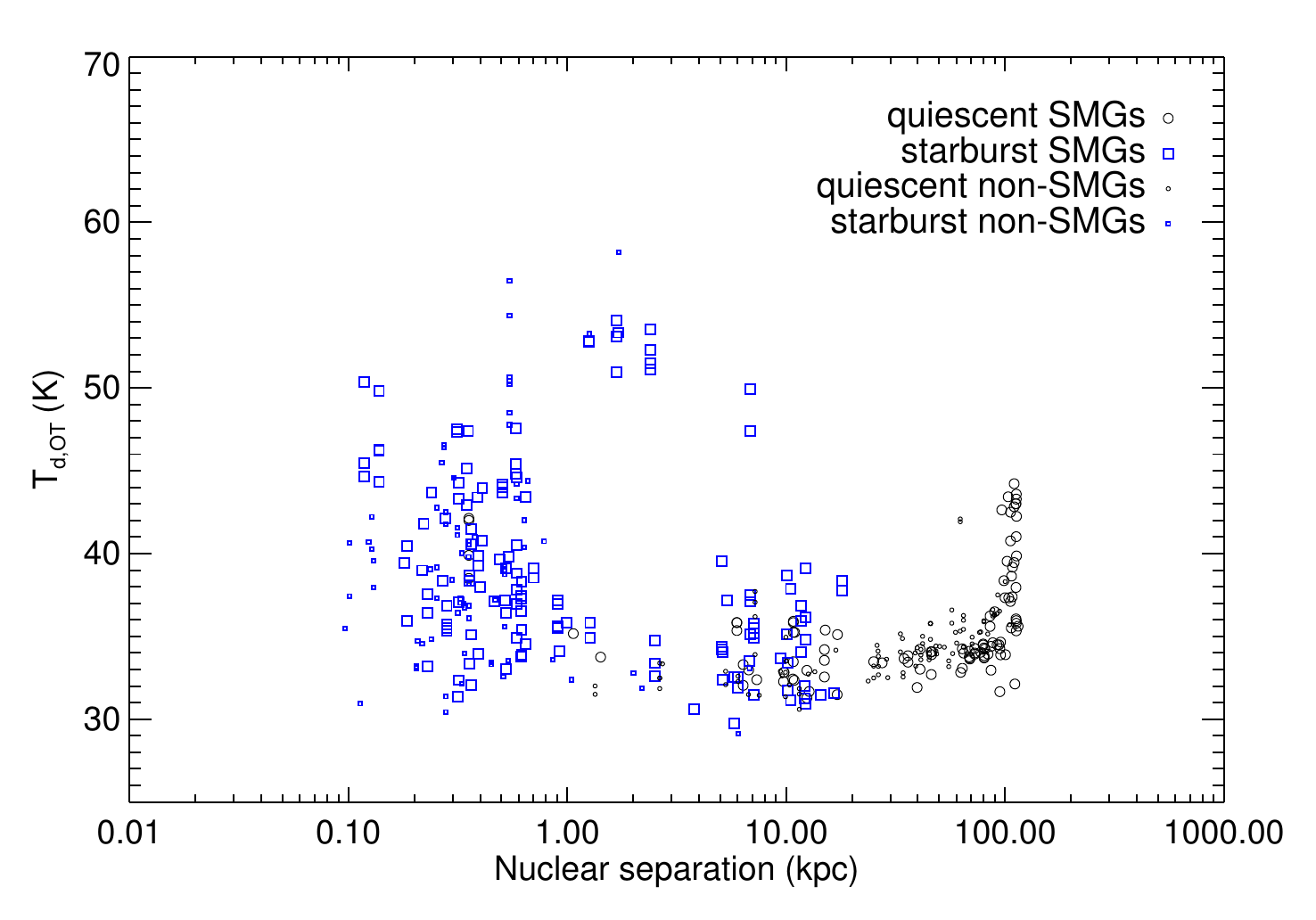}
\\
\includegraphics[width={0.90\columnwidth}]{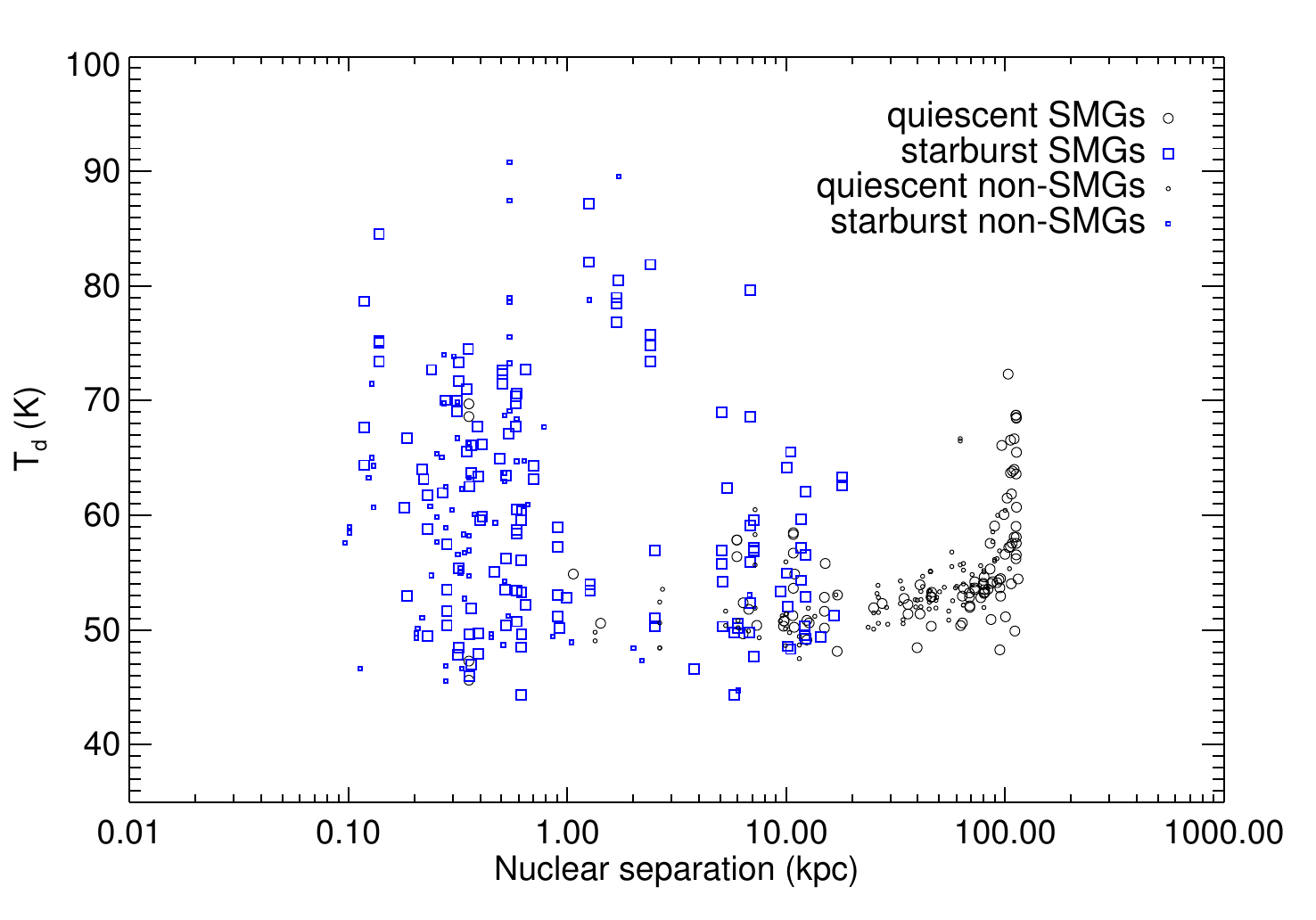}
\\
\includegraphics[width={0.90\columnwidth}]{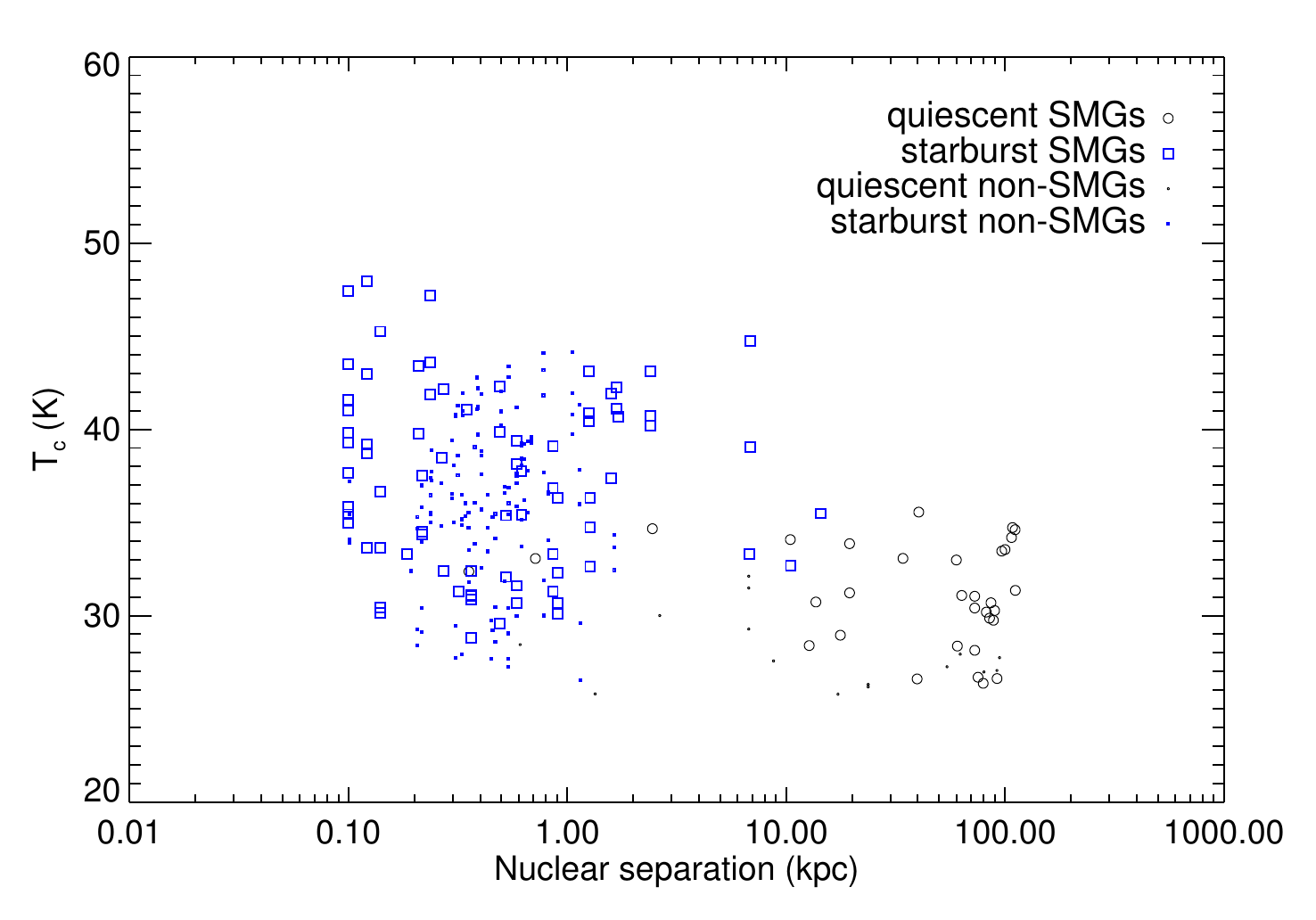}
\caption{Effective dust temperature derived from fitting the single-$T$ OT modified blackbody (Equation \ref{eq:mod_BB_ot}; top),
full modified blackbody (Equation \ref{eq:mod_BB}; middle), and power-law $T$-distribution model (bottom)
to the PACS+SPIRE+SCUBA+AzTEC integrated photometry versus nuclear separation
for snapshots where the BHs have not coalesced. $\tdust$ is anti-correlated with nuclear separation,
but there is a large scatter in effective temperature at a given nuclear separation.
The objects with the highest effective dust temperatures ($\tdustot \ga$ 35 K, $\tdust \ga$ 55 K, and $\tc \ga 36$ K)
are almost exclusively late-stage merger-induced starbursts with $d_{\rm BH} \la 10$ kpc.
The simulations with high $\tdust$ at $\dbh \sim 100$ kpc are those where $\lir$ -- and thus $\tdust$ -- is still high (but instantaneous SFR is not)
because of the burst of star formation that can occur at first passage.}
\label{fig:T_d_vs_d_bh}
\end{figure}

Figs. \ref{fig:L-T_ot}, \ref{fig:L-T}, and \ref{fig:L-T_pl} all show that the $\tdust-L_{\rm IR}$ plot is an excellent way to select starburst SMGs from the general population.
In all three figures there is a clear correlation between $T_{d}$ and $L_{\rm IR}$, which agrees with observations of both local
and high-redshift ULIRGs (e.g., \citealt{Kovacs:2006,Magnelli:2010,Amblard:2010,Chapman:2010,Hwang:2010dust_T_evolution,Magdis:2010dust_T}; M12).
Though there is some overlap between the quiescently star-forming and starburst sub-populations, the most luminous, hottest SMGs are almost exclusively starbursts.
Note that both the $\tdust-L_{\rm IR}$ correlation and the separation between the populations are independent of the fitting method used,
though the specific temperature values above which there are no quiescently star-forming galaxies differ (as expected because of the systematic
difference in temperatures yielded by the two methods). Thus our simulations make the clear, robust prediction that the most
luminous galaxies will have the hottest SEDs and will typically be late-stage merger-induced starbursts.

In all cases there is a sub-population of hot-dust ULIRGs which have relatively high $\tdust$ for a given IR luminosity. Such galaxies are not present in the SMG sub-population
because of the bias of the SMG selection. The SMG selection bias results in an apparent relatively tight correlation between $\tdust$ and $\lir$, so an $\lir$ cut is roughly as
effective as a $\tdust$ cut at selecting a subset of starburst SMGs from the general SMG population. When the hot-dust ULIRGs are included the scatter in the $\tdust-\lir$
relation for $\lir \la 10^{13} \lsun$ is increased significantly.
For $\lir \la 10^{13} \lsun$ a $\tdust$ cut can select starbursts whereas an $\lir$ cut cannot.
Thus if one wishes to select starbursts from a given galaxy population a $\tdust$ cut is preferred because it can isolate a larger subsample of starbursts with a wider
range in $\lir$.

Fig. \ref{fig:T_d_vs_d_bh} shows $\tdust$ derived from fitting the single-$T$ OT modified blackbody form (top), full form of the single-$T$ modified blackbody (middle), and
power-law $T$-distribution model (bottom)
to the PACS, SPIRE, SCUBA, and AzTEC photometry (with some PACS data excluded for each form, as described above)
versus separation of the central BHs (aka nuclear separation; $\dbh$). The nuclear
separation serves as a proxy for the merger stage, but it is important to keep in mind that it does not decrease monotonically as the merger progresses.
The starburst galaxies have systematically lower $\dbh$ than the quiescently star-forming galaxies because the tidal torques which drive the
starburst are strongest at final coalescence of the two discs. The typical $\tdust$ values increase with decreasing $\dbh$, though there is large scatter at
a given $\dbh$, especially for $\dbh \la 10$ kpc. This occurs because, for a given simulation, $\lir$ is anti-correlated with $\dbh$ (because the starburst is
strongest at low nuclear separations), and, as seen above, the most luminous galaxies are also
the hottest. Interestingly, for a given nuclear separation the SMGs and non-SMGs span a similar range in $\tdust$. The simulated mergers with $\dbh \sim 100$ kpc
and relatively high $\tdust$ are those observed shortly after first passage. At this time $\lir$ is high and the starburst mode is important even though the galaxies do
not meet the definition of starburst because the instantaneous SFR has dropped.

\subsubsection{Comparison of fitting forms} \label{S:sed_fitting}

\begin{figure}
\centering
\plottwo{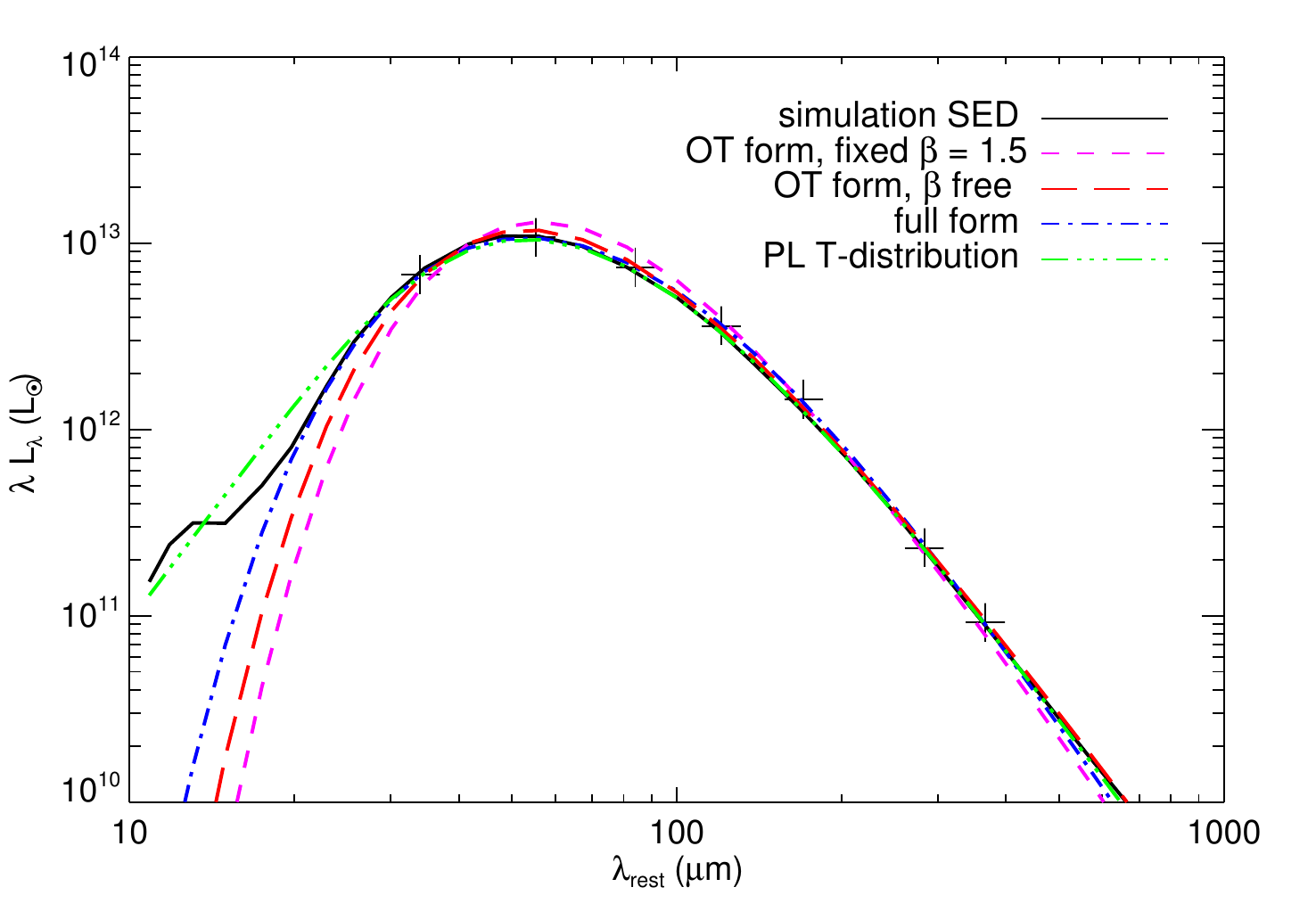}{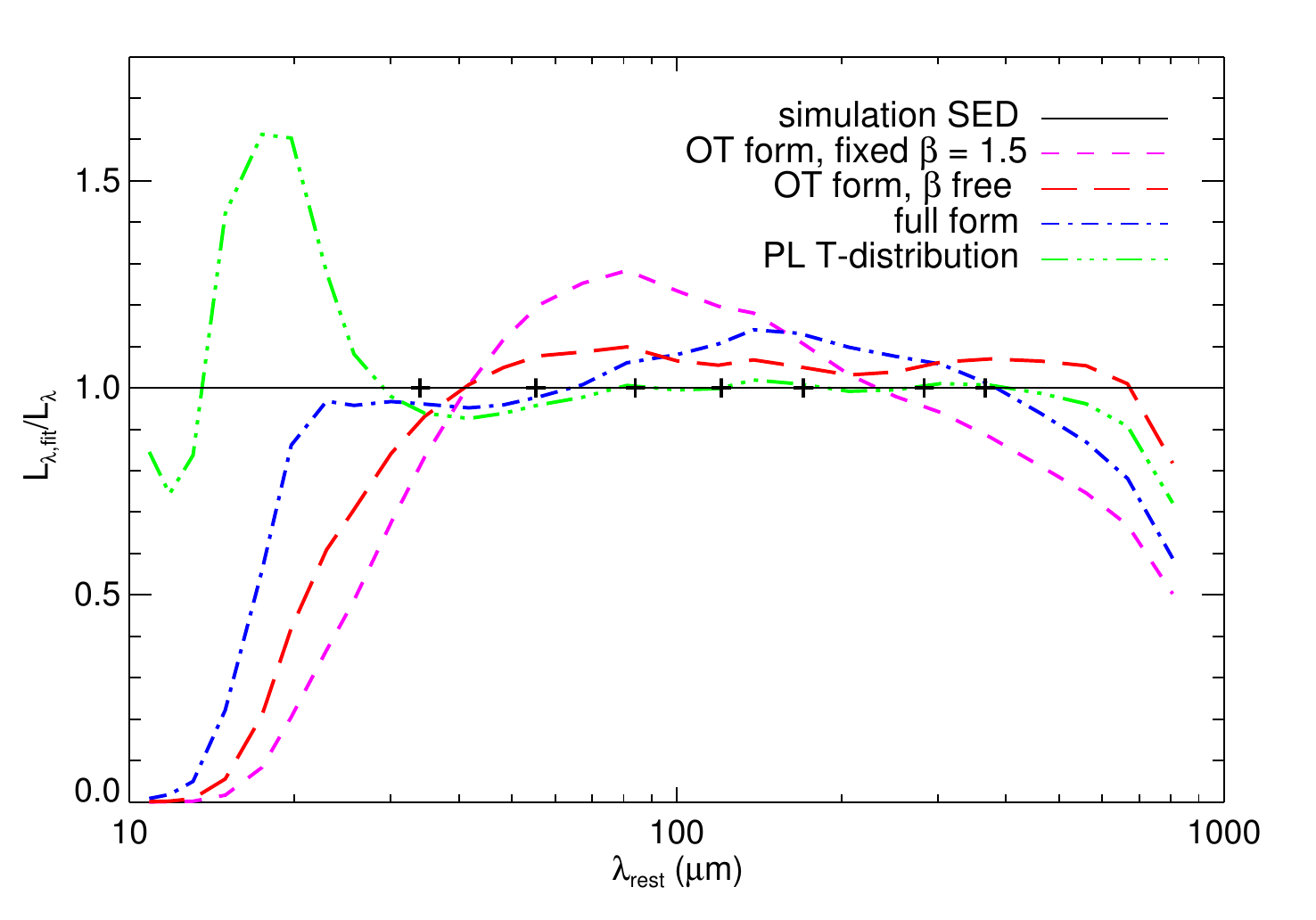}
\caption{Top: The black line is an example rest-frame FIR SED, $\lambda L_{\lambda}$ $(\lsun)$ versus $\lambda_{\rm rest}$ ($\micron$), of a simulated starburst.
The crosses are the PACS 100- and 160-
$\micron$, SCUBA 250-, 350-, and 500-$\micron$, SCUBA 850-$\micron$, and AzTEC-1.1 mm photometry calculated from the simulation SED. The other lines are
best fits to the photometric points for different fitting forms: the single-$T$, OT modified blackbody (Equation \ref{eq:mod_BB_ot}) with fixed $\beta = 1.5$
(magenta dashed), the single-$T$ OT modified blackbody with $\beta$ allowed to vary (red long-dashed),
the full form of the single-$T$ modified blackbody (Equation \ref{eq:mod_BB}; blue dash dot),
and the power-law $T$-distribution (\citealt{Kovacs:2010}; green dash dot dot dot). Bottom: Ratio of the fitted $L_{\lambda}$ to the actual $L_{\lambda}$
versus $\lambda_{\rm rest}$ $(\micron)$. The crosses overlaid on the black solid line, which denotes $L_{\lambda, \mathrm{fit}} = L_{\lambda}$,
show the wavelengths of the photometric points used for the fits.
The values of $\tdust$ and $\beta$ for the best-fitting models are given in Table \ref{tab:sed_fit}.
All forms except the single-$T$ OT form with fixed $\beta$ recover the photometry within the 10 per cent uncertainty.
However, the derived dust temperature and $\beta$ depend strongly on the fitting method, so the model
parameters should not be interpreted physically even though the models adequately describe the data.}
\label{fig:sed_fit}
\end{figure}

\ctable[
    caption	=	{$\tdust$ and $\beta$ for the best-fitting models shown in Fig. \ref{fig:sed_fit} \label{tab:sed_fit}},
    			center
  ]{lcc}{  													
  }{
  														\FL
    Model 							& $\tdust$		& $\beta$		\ML
    Single-$T$ OT form, fixed $\beta$		& 47 K		& 1.5			\NN
    Single-$T$ OT form				& 53			& 1.1			\NN
    Single-$T$ full form				& 69			& 1.5			\NN
    Power-law $T$-distribution			& 49			& 1.2			\LL
  }

In the top panel of Fig. \ref{fig:sed_fit} we show the rest-frame SED of one of the simulated starbursts viewed from a single viewing angle. The over-plotted
data points are the simulated photometry as described in the caption.
The four lines are fits to the photometry using the models discussed above. The bottom panel shows the ratio of the model SED derived from fitting the photometry
to the actual SED for each of the fitting methods. The values of $\tdust$ and $\beta$ for the best-fitting models are given in Table \ref{tab:sed_fit}.

All forms except the single-$T$ OT modified blackbody with fixed $\beta = 1.5$ are able to recover the simulated photometry to within
$\sim 10$ per cent, which is the level of uncertainty assumed when performing the $\chi^2$ minimisation. The single-$T$ OT form
with fixed $\beta$ can only recover the photometry to within $\sim 30$ per cent.
Though the more complicated forms are successful at recovering the photometry used for the fitting, they
have varying levels of success describing the SED beyond the wavelengths spanned by the photometry.
As explained above, we expect the OT modified blackbody to under-predict the
SED on the Wien side of the SED. Indeed, this model under-predicts the SED shortward of the 100-$\micron$ point, and the
under-prediction is more severe when $\beta$ is fixed. The full form of the modified
blackbody fares better, but it also under-predicts the SED for rest-frame wavelength $\lambda_{\rm rest} \la 20~\micron$. 
The power-law temperature distribution model fares best at the shortest wavelengths, but it over-predicts the SED at $\lambda_{\rm rest} \sim 15-25~\micron$.
At the longest wavelengths the single-$T$ OT and power-law $T$-distribution models are most accurate because they have relatively low $\beta$ values and thus less
steeply declining SEDs.

Table \ref{tab:sed_fit} shows that the derived parameters $\tdust$ and $\beta$, which are often interpreted as a physical dust temperature and the intrinsic power-law index of the emissivity of the
dust grains in the FIR, vary for the different fitting forms. Since observed galaxies typically have $20 \la \tdust \la 80$ K and $1 \la \beta \la 2$, the variation among the best-fitting values
is very significant. Consequently, it is difficult to interpret the fitted parameters physically, as clearly the intrinsic properties
of the dust do not vary with the method used to fit the SED.\footnote{Note that the issues discussed here are independent of observational noise, as we have added
none to our simulated SEDs. Observational noise further complicates interpretation of the derived parameters \citep{Shetty:2009a,Kelly:2012}.}
If the fitted effective dust temperatures have a physical meaning, they may correspond to different
physical temperatures (e.g., the single-$T$ modified blackbody may recover the luminosity-weighted dust temperature whereas the power-law model may better
recover the mass-weighted temperature). None of the fitted $\beta$ values recover the intrinsic $\beta$ of the dust, which is $\sim 2$ for the
dust model we use.
This is not unexpected, because a distribution of physical dust temperatures will change the slope of the SED and thus the $\beta$ in a single-$T$ model;
non-negligible optical depths in the IR further complicate the picture.

Clearly it is necessary to determine how the fitted
parameters relate to intrinsic properties of the dust, but we defer further exploration of this complex topic to future work. We only wish to stress
that it may be necessary to use forms more sophisticated than the single-$T$ OT modified blackbody to fit IR SEDs and that the parameters derived
from the fits should not be interpreted physically. Instead, the models should be thought of as useful ways to encapsulate the data with a few parameters and
thus compare galaxy SEDs in a simple way by comparing the fitted parameters. Put another way, the $\tdust-\lir$ plot still contains useful information
about SED variation even if $\tdust$ is not a physical dust temperature, and differences in $\tdust$ amongst galaxies reflect real differences in the galaxies'
SEDs.

\subsection{Star formation efficiency} \label{S:SFE}

\begin{figure}
\centering
\plottwo{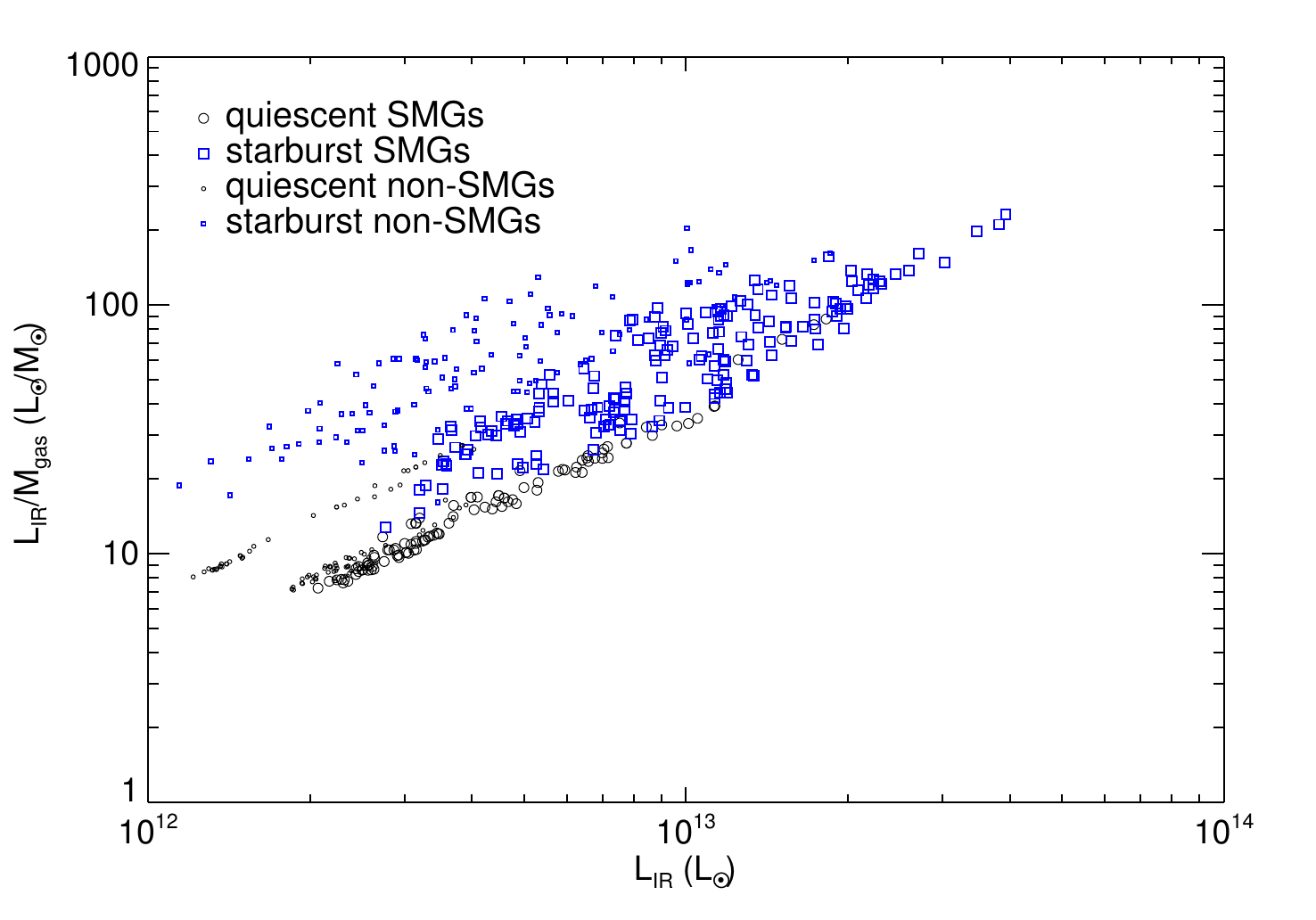}{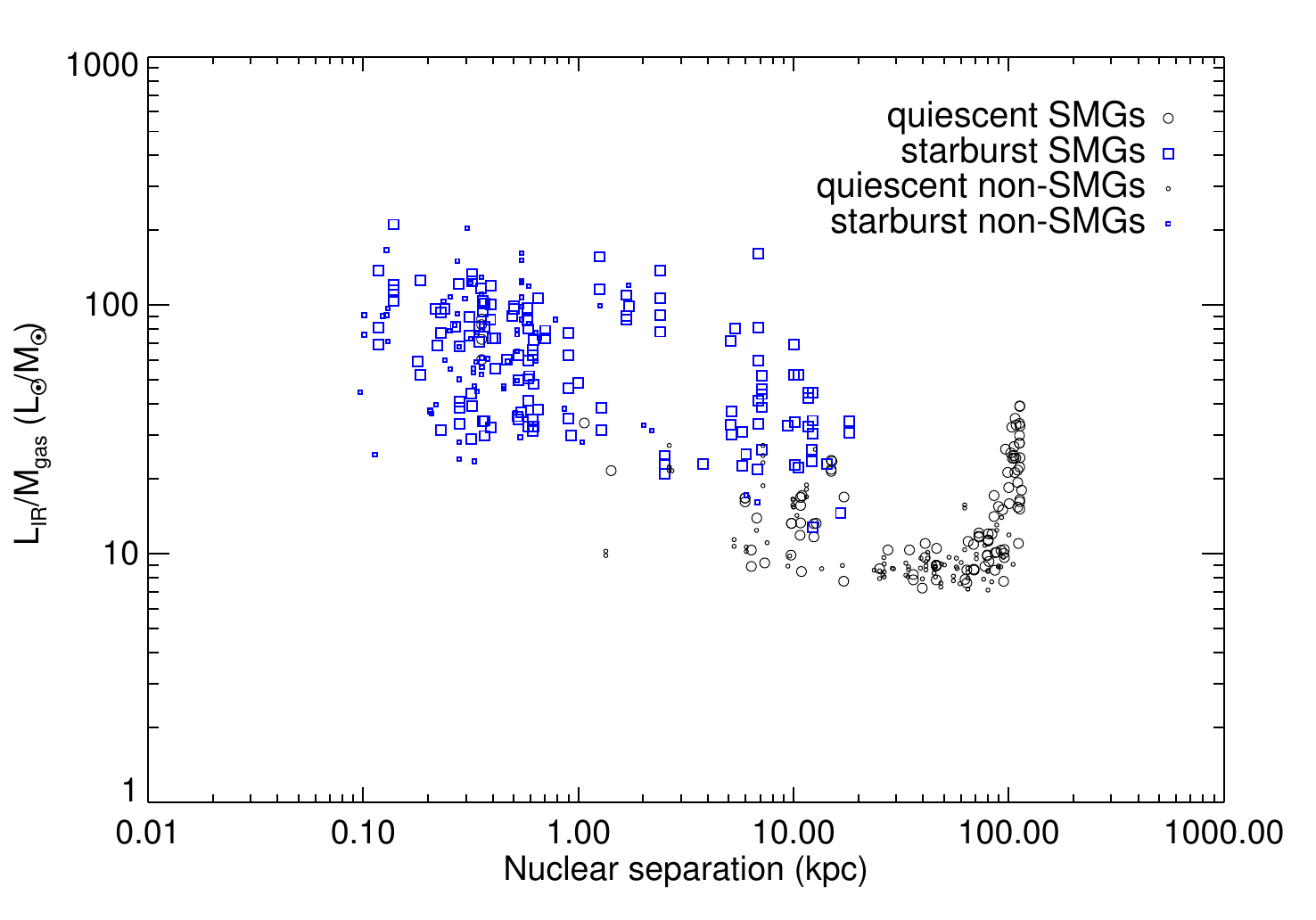}
\caption{`Star formation efficiency' SFE $\equiv \lir/\mgas$ ($\lsun/\msun$) versus $\lir$ ($\lsun$; top) and nuclear separation (kpc; bottom). SFE and $\lir$ are
totals for the system in order to mimic the effects of blending in the FIR and (sub)mm. At fixed $\lir$ the SFE
is characteristically higher by a factor of a few for starburst SMGs than for quiescently star-forming SMGs, and the starbursts (quiescently star-forming
galaxies) missed by the SMG selection have SFE greater than the starburst (quiescently star-forming) SMGs. The spread in SFE at fixed $\lir$
between the most extreme starburst non-SMGs and the quiescently star-forming SMGs can be as much as $\sim 10 \times$.
The starbursts near coalescence have the highest SFE. The objects with high SFE at $\dbh \sim 100$ kpc are the same as the high-$\tdust$ objects
with $\dbh \sim 100$ kpc, mergers observed shortly after first passage when $\lir$ is still high but instantaneous SFR is not.}
\label{fig:SFE}
\end{figure}

\ctable[
    caption	=	{Median SFE values ($\lsun/\msun$) \label{tab:SFE}},
    			center
  ]{lcc}{
  }{
      															\FL
      Galaxy type			& $\lir < 10^{13} \lsun$	& $\lir \ge 10^{13} \lsun$	\ML
      Quiescent SMGs		&	13.6				& 60.3				\NN
      Starburst SMGs		&	37.1				& 91.2				\NN
      Quiescent non-SMGs	&	9.4				& --					\NN
      Starburst non-SMGs	&	51.2				& 123.4				\LL
}

Since starbursts form stars much more efficiently than quiescently star-forming galaxies, one should be able to distinguish between them via some
observational indicator of star formation efficiency. To be consistent with the literature \citep[e.g.,][]{Daddi:2010,Genzel:2010} we define `star formation efficiency' as
SFE $\equiv \lir/\mgas$.
Note, however, that $\lir$ does not necessarily track the instantaneous SFR because, in addition to recently formed stars, older stars and
AGN can also contribute to $\lir$ (see, e.g., section 2.5 of \citealt{Kennicutt:1998} and section 3.1 of H11 for details), so this should be considered only an approximate indicator of star formation efficiency.
Furthermore, inferring the total molecular gas mass from CO observations is notoriously difficult, as the CO--H$_2$ conversion factor
$X_{\rm CO} = N_{\rm H2}/L'_{\rm CO}$ depends on the giant molecular cloud surface density and the kinetic temperature and velocity dispersion within clouds
\citep{Narayanan:2011X_CO_I,Narayanan:2011X_CO_II,Shetty:2011X_CO_I,Shetty:2011X_CO_II,Papadopoulos:2012}. As a result, $X_{\rm CO}$ is
expected to be a factor of $\sim2-10\times$ lower in starbursts than in disc galaxies \citep{Narayanan:2011X_CO_I}.
The uncertainty surrounding $X_{\rm CO}$ complicates efforts to determine how much the SFE of starbursts and quiescently star-forming galaxies differs
\citep[e.g.,][]{Papadopoulos:2012}.
It would be best for us to predict the CO line luminosity for our simulated galaxies, as has been done by, e.g.,
\citet{Narayanan:2009}, and, ideally, to self-consistently track formation and destruction of molecular gas \citep[see, e.g.,][]{Robertson:2008mol_gas}.
However, doing so requires introduction of another code -- and the corresponding complexities and uncertainties -- in addition to the two employed,
so we feel this is best left to future work (though we briefly discuss how the molecular gas emission might be used as a diagnostic in Section \ref{S:other_diagnostics}).
Thus when calculating SFE here we use the total gas mass instead of the molecular gas mass.

In Fig. \ref{fig:SFE} we plot SFE versus $\lir$ (top) and nuclear separation (bottom); the median values are given in Table \ref{tab:SFE}.
At a given $\lir$, the starburst SMGs have SFE
up to 5$\times$ greater than that of the quiescently star-forming galaxies. The starbursts not selected as SMGs have the highest SFE, whereas the
quiescently star-forming SMGs have the lowest values; the discrepancy can be as much as $\sim 10 \times$.
The bottom panel of Fig. \ref{fig:SFE} demonstrates that for both the SMGs and non-SMGs the SFE increases as the merger advances and is highest for mergers nearest coalescence.
At a given $\dbh$ the SMGs and non-SMGs span a similar range in SFE. The objects with high SFE at $\dbh \sim 100$ kpc are mergers observed shortly after first passage. For
these galaxies $\lir$ is still high, causing high SFE, but instantaneous SFR is not, so they do not meet the definition of starburst.

The simulations qualitatively reproduce the systematic offset in SFE between quiescently star-forming galaxies and starbursts shown in fig. 1
of \citet{Daddi:2010}. However, for a given $\lir$ the SFE values of the simulated starbursts are significantly ($\sim 5 \times$) lower than the observed values.
Part of this is because our measure of the SFE in the simulations uses the total gas mass, not the molecular gas mass.
In the simulations at the peak of the starburst, cold gas within 5 kpc of the centre, which we take as a rough approximation for molecular gas that would be probed by
observations \citep[see][]{Narayanan:2009}, is typically less than half of the total gas mass even though it accounts for effectively all of the star formation.
Thus if we were to calculate SFE for our simulated starbursts using molecular gas mass rather than total gas mass the values for the simulations would be a factor
of $\sim 2-3 \times$ higher, which would account for a large part of the discrepancy between the simulated and observed starbursts.

The SFE of the simulated starburst SMGs is only a factor of $\sim2-3\times$ greater than that of the simulated quiescently star-forming galaxies, whereas the observed
difference is $\sim4-10\times$. For the starbursts nearest coalescence, however, the difference can be as great as $10\times$ (see bottom panel of Fig. \ref{fig:SFE}).
One possible reason the SFE discrepancy is lower than observed is that, because of the setup of our simulations,
the pre-burst, quiescently star-forming discs have systematically higher gas fractions than those discs near coalescence. Since the star formation law implemented in the
simulations has a non-linear dependence on gas density the SFE will increase with gas fraction.
Despite the possible discrepancies described above, the simulations make the robust prediction that those systems with largest SFE at a given $\lir$
will be merger-induced starbursts.

\subsection{IR excess} \label{S:obscuration}

\begin{figure}
\centering
\plottwo{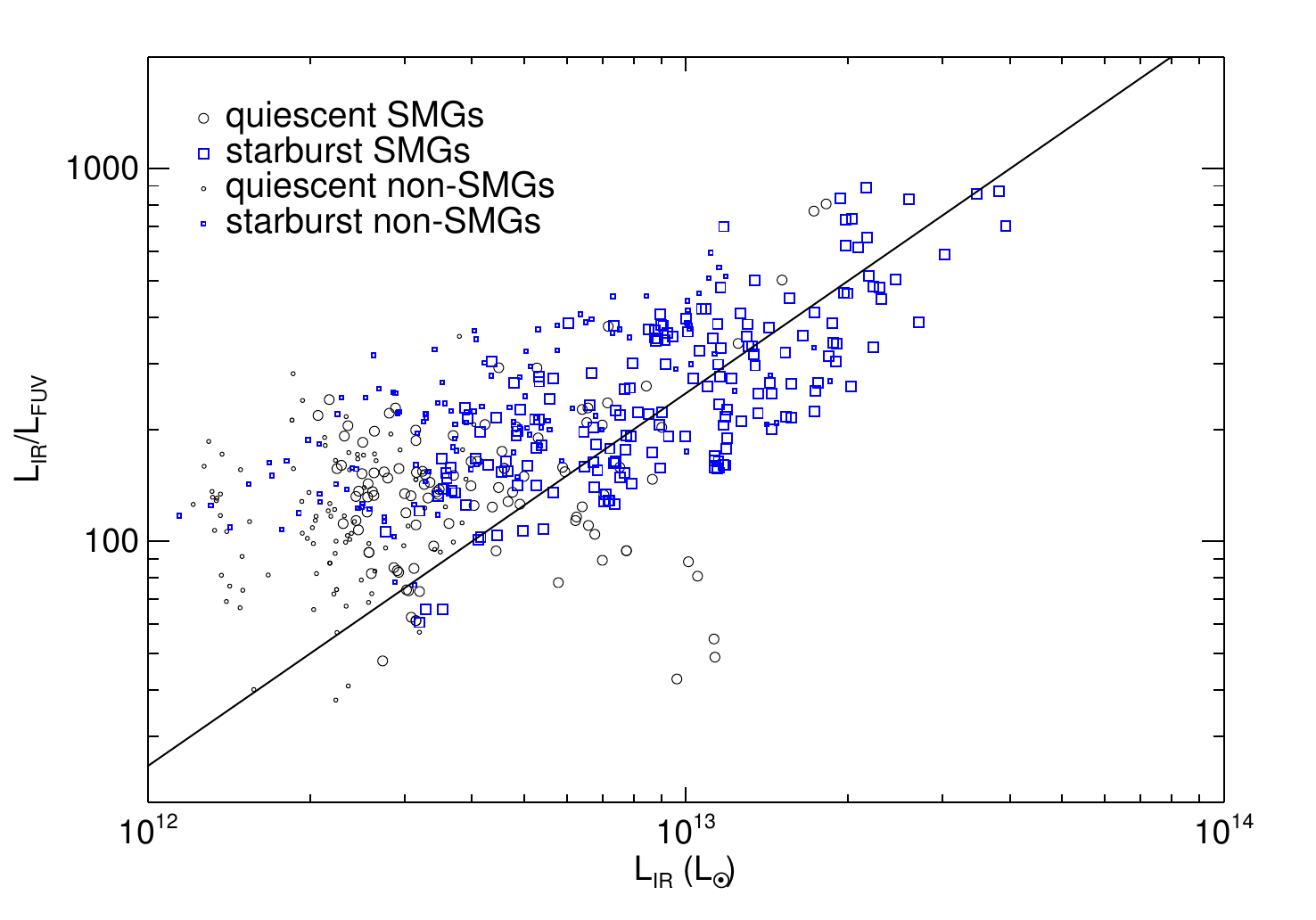}{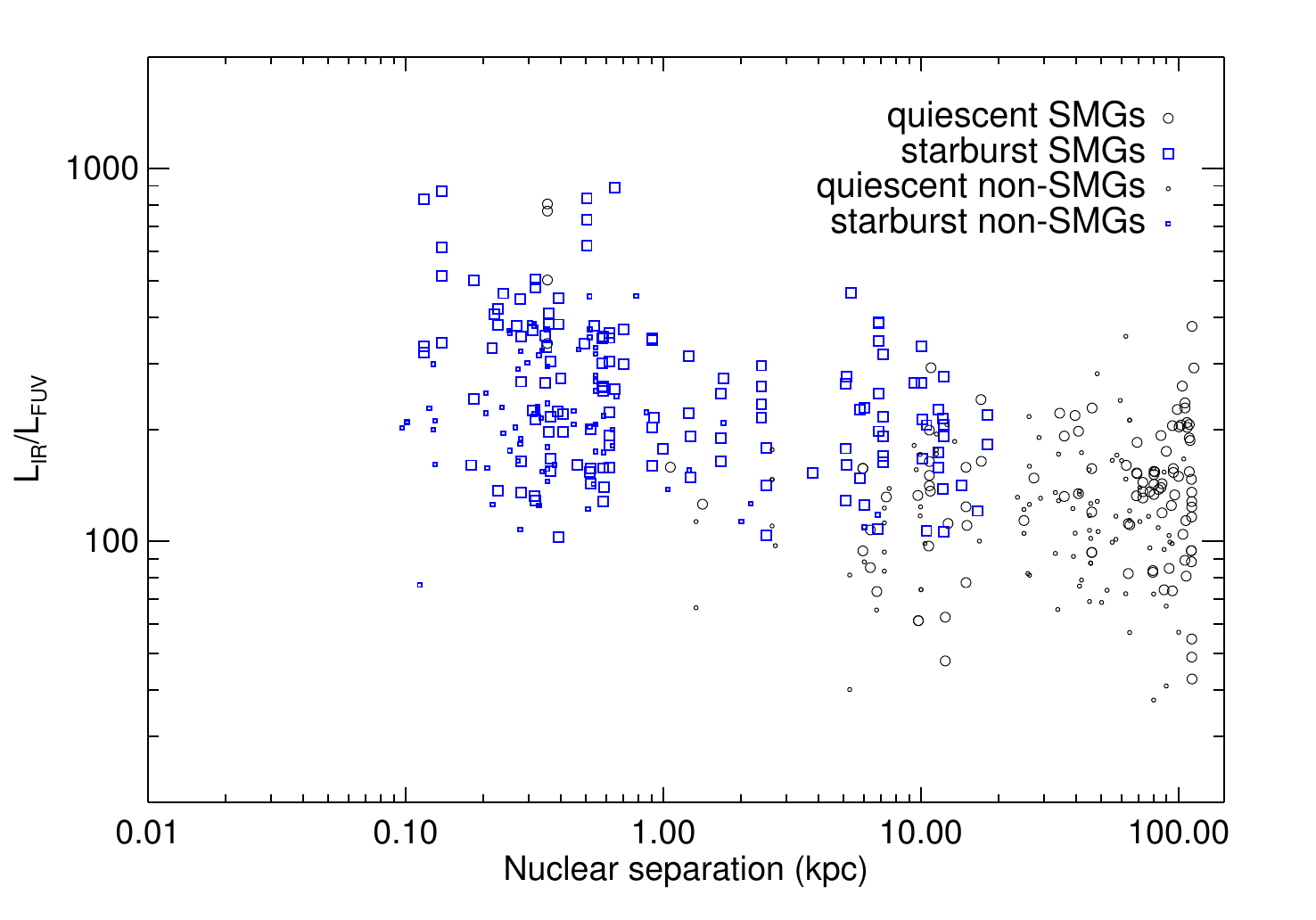}
\caption{IR excess (IRX $\equiv \lir/\lfuv$) versus $\lir$ ($\lsun$; top) and nuclear separation (kpc; bottom). Both IRX and $\lir$ are calculated for the total
system. For the starbursts, IRX is correlated with $\lir$ and
anti-correlated with nuclear separation. The starbursts are typically more obscured (have higher IRX) than the quiescently star-forming galaxies. However, for
fixed $\lir$ the starbursts and quiescently star-forming discs have similar IRX values, so the reason starbursts typically have higher attenuation is because
they are typically more luminous than the quiescently star-forming discs. Interestingly, the correlation is reasonably well-approximated by constant
$\lfuv = 4 \times 10^{10} \lsun$ (solid line in the top panel) at the bright end.}
\label{fig:obscuration}
\end{figure}

\ctable[
    caption	=	{Median IRX values \label{tab:IRX}},
    			center
  ]{lcc}{
  }{
      															\FL
      Galaxy type			& $\lir < 10^{13} \lsun$	& $\lir \ge 10^{13} \lsun$	\ML
      Quiescent SMGs		&	144.9			& 340.0				\NN
      Starburst SMGs		&	185.4			& 332.3				\NN
      Quiescent non-SMGs	&	120.1			& --					\NN
      Starburst non-SMGs	&	199.9			& 373.3				\LL
}

In Fig. \ref{fig:obscuration} we plot the total IR luminosity divided by the rest-frame far-UV luminosity, $\lir/\lfuv$ --
referred to as the IR excess (IRX) -- versus $\lir$ (top) and nuclear separation (bottom); the median values are given
in Table \ref{tab:IRX}. Both IRX and $\lir$ are calculated for the total system. The IRX serves as a measure of the level
of obscuration of a galaxy.
IRX increases with $\lir$, as has previously been both observed \citep[e.g.,][]{Wang:1996,Buat:1998,Buat:1999,Buat:2005,Buat:2007,Buat:2009,
Adelberger:2000,Hopkins:2001,Bell:2003,Reddy:2010} and demonstrated by simulations \citep{Jonsson:2006}.
The quiescently star-forming galaxies tend to have lower IRX than the starbursts, primarily because the
starbursts are typically more luminous. At a given $\lir$, the starbursts and quiescently star-forming discs have very similar IRX.
\citet{Jonsson:2006} have previously demonstrated this result using similar simulations, and they argue that the correlation arises because both SFR
and dust optical depth correlate with density.
The bottom panel shows that IRX increases as nuclear separation decreases. This adds further evidence in support of the arguments of \citet{Jonsson:2006}:
The galaxies are more compact at coalescence than during the pre-burst, infalling-disc stage. The resulting higher densities lead to higher SFR and thus
greater $\lir$.
Furthermore, the stars formed in the starburst, which dominate the luminosity, are centrally concentrated and thus typically more obscured
than stars distributed throughout the initial discs. Thus $\lir$ and IRX both increase as nuclear separation decreases.
This finding is consistent with that from studies of AGN obscuration using similar simulations
\citep{Hopkins:2005quasar_evolution,Hopkins:2005quasar_lifetimes,Hopkins:2006unified_model}.

The solid line in the top panel of Fig. \ref{fig:obscuration} is the relation given by constant $\lfuv = 4 \times 10^{10} \lsun$.
Note that, assuming this value, $\lfuv$ is less than one per cent of the bolometric luminosity. At the bright end this relation
approximates that of the simulated galaxies reasonably well, suggesting that the light observed in the rest-frame FUV is decoupled from the bolometric luminosity.
This what we expect when the luminosity of a galaxy is dominated by a deeply dust-enshrouded starburst and
the only light observed in the FUV is from stars located outside the heavily obscured nuclear region \citep[see also][]{Jonsson:2006}.
At lower luminosities $\lfuv$ decreases as $\lir$ decreases, causing the simulated galaxies to lie above the $\lfuv = 4 \times 10^{10} \lsun$ relation.
In particular, the quiescently star-forming galaxies tend to lie above the relation because they are much less obscured than the starbursts.

\subsection{SFR--$\mstar$ relation} \label{S:SFR-Mstar}

\begin{figure}
\centering
\plotone{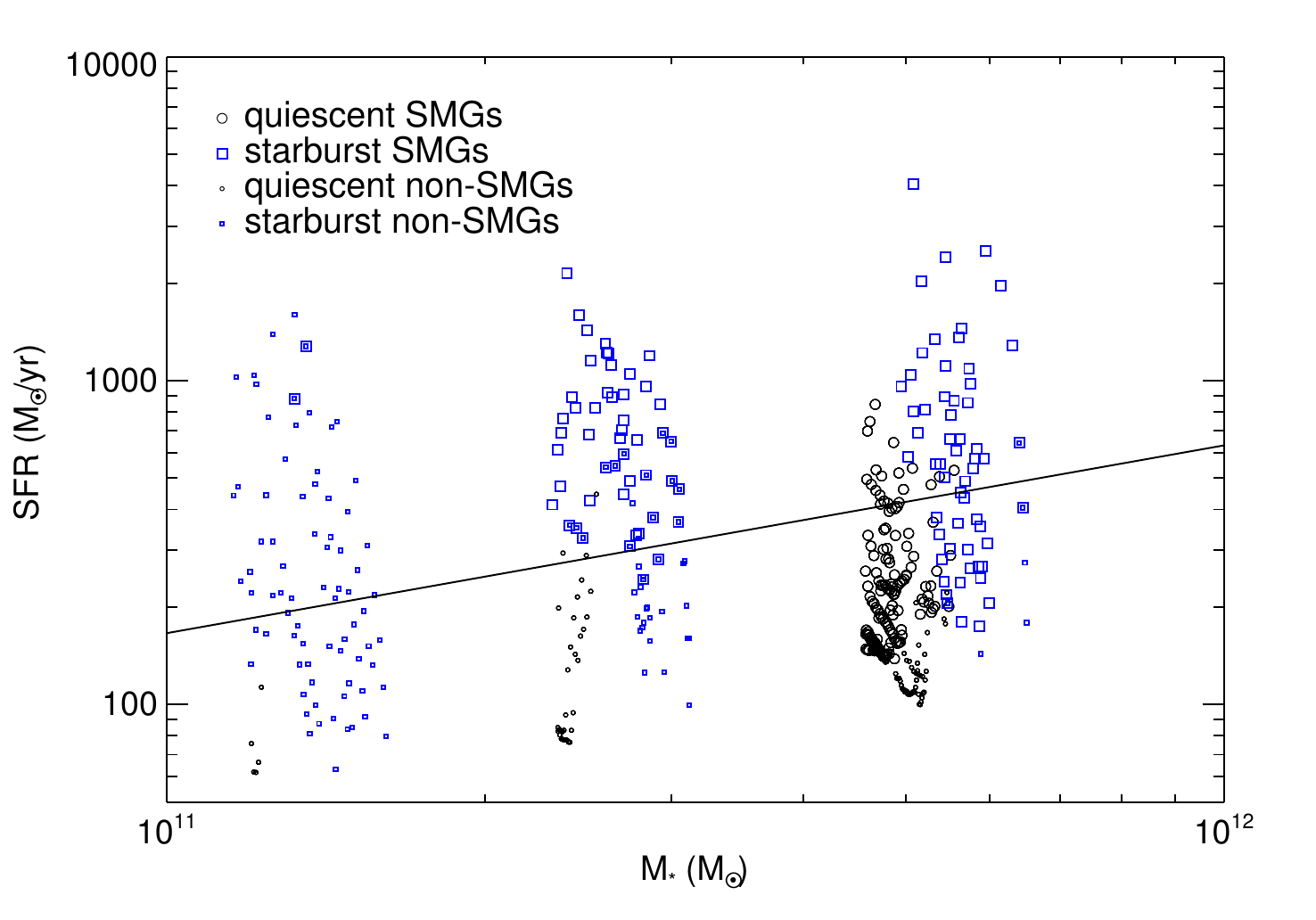}
\caption{SFR $(\msunperyr)$ versus $\mstar$ ($\msun$), where both are totals for the system. The solid line is the relation for the $z = 2.0-2.5$ bin
of \citet{Karim:2011}. Simulated galaxies above the observed relation are almost exclusively starbursts, and the starbursts typically
lie above the relation.}
\label{fig:SFR-M*}
\end{figure}

As discussed in Section \ref{S:SF_modes}, if the tight SFR--$\mstar$ relation observed is set by the gas supply rate then galaxies that lie above the relation
must be undergoing transient events that temporarily boost their SFR above what can be sustained over long time periods. Major
mergers are one type of event that can cause galaxies to move above the relation. We plot the SFR--$\mstar$ relation for our
simulated galaxies in Fig. \ref{fig:SFR-M*} along with the observed relation for $z = 2.0-2.5$ from \citet[][solid line]{Karim:2011}.
Almost all objects above the observed relation are starbursts, and a large fraction of the starbursts are above the relation.
For a given $\mstar$ the starbursts can have SFR values $\ga 10\times$ those of the quiescently star-forming galaxies. 

Note that some of the simulated galaxies, including starbursts, lie significantly below the relation. This is partially caused by the idealised setup of the simulations:
Since there is no cosmological gas accretion some of the simulated galaxies have gas fractions lower than what is expected from observations and cosmological simulations.
For fixed galaxy size and total mass, the observed KS relation implies that SFR scales super-linearly with gas fraction, so simulated galaxies with gas fractions
lower than real galaxies will have significantly lower SFR than observed. Note also that the pre-burst discs have systematically higher gas fraction because the
gas fraction decreases monotonically with time in the simulations. This biases the quiescently star-forming galaxies' SFR high relative to the starbursts,
so the magnitude of the difference between starbursts and quiescently star-forming discs' SFR shown in Fig. \ref{fig:SFR-M*} should be taken as a
lower limit. Furthermore, discrepancies in the SFR--$\mstar$ relation of the simulations and that observed may occur because the SFR derived using a given
diagnostic is not equivalent to the instantaneous SFR of the simulated galaxies, as even the observed relations can differ depending on what SFR diagnostic
is used to derive them.

Observationally, whether SMGs lie on the SFR--$\mstar$ relation depends on the measured $\lir$ used to infer the SFR
and the inferred $\mstar$. The latter is especially difficult to determine: different authors have inferred masses differing by a factor
of $\sim 6 \times$ for the same SMGs (\citealt*{Michalowski:2010masses,Michalowski:2010production}; \citealt{Michalowski:2011,Hainline:2011}).
If the \citet{Michalowski:2011} masses are used, SMGs lie much closer to the SFR--$\mstar$ relation than they do if the \citet{Hainline:2011} masses are used.
However, in both cases a subset of SMGs are significant outliers from the relation; this conclusion is consistent with our claim that the SMG population
is a mix of both quiescently star-forming and starburst galaxies.

\section{Discussion} \label{S:discussion}

\subsection{The need to distinguish star formation modes}

If one wishes to understand star formation it is crucial to look beyond the local universe, because the SFR density of the universe
was greatest at $z \sim 2-3$ \citep[e.g.,][]{Madau:1996,Steidel:1996,Hopkins_A:2004,Hopkins_A:2006,Hopkins:2010IR_LF,Karim:2011,Magnelli:2011}.
Furthermore, the bulk of the star formation at those
redshifts was obscured \citep[e.g.,][]{Bouwens:2011,Magnelli:2011}, so studying IR-luminous galaxies at those redshifts is crucial.
Unfortunately, galaxies become fainter and physical resolution poorer as one moves from $z \sim 0$ to higher redshift,
so observations of high-redshift galaxies are significantly less detailed than for local galaxies.
It is thus tempting to use wisdom gleaned from detailed observations of local galaxies to guide the interpretation of observations of high-redshift
galaxies. This is perfectly acceptable if the only difference between local galaxies and those at $z \sim 2-3$ is that the latter are farther away.
However, this is clearly not the case, so one must apply local-universe-derived wisdom with caution.

Assuming what is true locally is also true at $z \sim 2-3$ can be problematic. For example, as discussed in Section \ref{S:SFE},
locally it seems that the CO--H$_2$ conversion factor $X_{\rm CO}$ differs for ULIRGs (i.e., merger-induced starbursts) and quiescently star-forming disc galaxies
\citep[e.g.,][]{Solomon:1997,Downes:1998}. If one wishes to, e.g., study possible evolution of the KS
law with redshift then, lacking other options, it is necessary to assume some CO--H$_2$ conversion factor for the high-redshift galaxy populations observed.
Choosing an appropriate $X_{\rm CO}$ requires determining whether the high-redshift galaxies are analogous to local merger-induced starbursts
or quiescently star-forming discs. 
For example, since it is commonly thought that SMGs are merger-driven starbursts, \citet{Daddi:2010} and \citet{Genzel:2010} use
the starburst $X_{\rm CO}$ value for SMGs. If, however, SMGs are a mix of quiescently star-forming, early-stage
mergers and late-stage, merger-induced starbursts, as we have argued in H11 and above, then a single $X_{\rm CO}$ value is not appropriate for the population.
In this case, use of the ULIRG $X_{\rm CO}$ value will artificially accentuate the apparent differences between SMGs and more typical galaxies at
high redshift.

These diagnostics can also be used to distinguish the quiescently star-forming sub-populations of SMGs (blended galaxy pairs and isolated discs)
from the starbursts in order to test the claims of our model and to understand the true nature of the population. We argue that galaxy pairs must contribute
significantly to the SMG population \citep[][Hayward et al., in preparation]{Hayward:2011num_cts_proc} because of the weak scaling
of submm flux with SFR in starbursts and the significantly longer duration of the galaxy-pair phase. However, given the modelling uncertainties
it is crucial to observationally determine the relative contributions.

Furthermore, if one wishes to understand which mode of star formation dominates the SFR density of the universe one must be able to separate
the modes. Even when one can clearly identify mergers (e.g., by the presence of tidal features) one cannot assume that those galaxies are dominated
by merger-induced star formation, as the SFR elevation caused by the mutual tidal torques is significant only near coalescence and, depending
on the gas content and bulge fraction of the progenitors, possibly first passage. (See \citealt{Hopkins:2010IR_LF} and \citealt{HH:2010} for further discussion of the
distinction between star formation in mergers and merger-induced star formation.) The problem is amplified at higher redshifts when mergers cannot
be easily identified.

\subsection{An observational roadmap to determine what star formation mode powers high-redshift ULIRGs}

Fortunately, the integrated SED of a galaxy contains much information about the star formation mode powering it,
so it is possible to use the diagnostics we have presented here to observationally disentangle what star formation mode dominates high-redshift ULIRGs.
This can be achieved by applying the diagnostics to the results of FIR and (sub)mm wide-field surveys. Since the diagnostics rely on
integrated data alone they are robust to blending in the FIR and (sub)mm, but the beam sizes at different wavelengths should be similar because otherwise
blending will be more severe at the wavelengths where resolution is poorer. The FIR and (sub)mm data are enough to use the $\tdust-L_{\rm IR}$ relation
as a diagnostic. Some of the other diagnostics require data at shorter wavelengths (to determine $\lfuv$ and $\mstar$), so care must be taken to include all
sources that contribute to the FIR emission, not just one component of a multiple component system. Keeping in mind the caveats discussed
above, one can also estimate gas masses and use the SFE as a diagnostic.

Once ALMA SMG surveys are available it will be simple to divide SMGs into single-component and multiple-component subclasses. This information
can be combined with $\lir$ and $\tdust$ values (preferably derived using the PL T-distribution model) to effectively divide the SMG population into the
various sub-populations: galaxy-pair SMGs will be resolved into two components, and those with smaller separations between components should typically
have higher $\lir$ and $\tdust$ values. The merger-induced starbursts and isolated discs will only have one component, but the former can be distinguished
by their higher $\tdust$ values.
Observations at rest-frame UV--NIR wavelengths can be used to apply the IRX and SFR--$\mstar$ diagnostics to further check the classification done
in the above manner, as starbursts will have higher values of IRX and lie above the SFR--$\mstar$ relation. If gas masses are available one can also use the SFE
as a diagnostic.

Spatial information beyond the number of components can further aid the classification. Though IFU spectrograph data provide high-resolution kinematics, high-redshift
ULIRGs can be optically thick well into the IR, so even rest-frame near-IR observations may not probe the central regions and thus must be interpreted with
caution.\footnote{\citet*{Rothberg:2010} have demonstrated that at $\lambda = 0.65$ \micron ~dust obscures the nuclear discs of young stars in local (U)LIRGs; since the galaxies
studied here are more obscured this effect should be more severe for high-redshift ULIRGs.}
Molecular gas emission, on the other hand, suffers significantly less dust attenuation, so (sub)mm interferometry with, e.g., ALMA can provide a direct view of the central
regions (but see \citealt{Papadopoulos:2010b,Papadopoulos:2010}).
The close pairs and isolated discs are most effectively distinguished via interferometry because the former will be resolved into two components with disc-like kinematics
(see, e.g., \citealt{Engel:2010}) whereas the latter will only show one disc. The merger-induced starbursts may show more disordered kinematics, but it is not always simple to
distinguish mergers shortly after coalescence from discs \citep[see][]{Robertson:2008}. Luckily, the starbursts can be distinguished from discs using
the diagnostics presented here.

By combining various data sets in the manner described above, it should be possible to separate high-redshift ULIRGs into starburst and quiescently
star-forming sub-populations. In addition, mergers can be roughly divided into widely separated pairs and close pairs, which are both quiescently
star-forming, and starbursts near coalescence. One can then more efficiently target sources for detailed follow-up, focusing on, e.g., different stages
in the merger process. Furthermore, by applying this technique to SMGs one can compare the sizes of the sub-populations to test our claim that
the SMG population is heterogeneous.

\subsection{Physical differences between SMGs and hot-dust ULIRGs} \label{S:smg_vs_non}

Throughout this work we have compared the properties of quiescently star-forming and starburst SMGs with simulated galaxies that would not be selected
by the SMG selection (`hot-dust ULIRGs'; \citealt{Chapman:2004,Chapman:2008,Casey:2009,Casey:2010,Magnelli:2010}; M12). At a given $\lir$
the hot-dust ULIRGs tend to have higher effective dust temperature than the SMGs; this is simply a consequence of the SMG selection because, for fixed $\beta$
and $\lir$, submm flux can be decreased only by increasing $\tdust$. Furthermore, at a given $\lir$
the hot-dust ULIRGs have higher SFE values than the SMGs, as suggested by \citet{Chapman:2008}, but the IRX values are similar.

The relative locations of the SMGs and hot-dust ULIRGs on the SFR-$\mstar$ diagram (Fig. \ref{fig:SFR-M*}) provide insight into the physical differences between these two
galaxy classes: The most massive galaxies are almost all selected as SMGs, regardless of whether they are quiescently star-forming or starbursts,
because they are luminous enough to have $S_{850} > 5$ mJy for any reasonable $\tdust$. At intermediate masses almost all of the starbursts are selected as SMGs, but
the quiescently star-forming galaxies are not. At the lowest masses simulated almost no galaxies are selected as SMGs. This reflects that fact that
galaxy mass is an important driver of the observed submm flux because it affects both the SFR and the dust mass (see also H11; \citealt{Michalowski:2011}).
The smaller galaxies can be very luminous in the IR if they are undergoing a strong starburst, but this results in a relatively hot SED (because $\tdust$ tends to
increase sharply during starbursts; H11) and causes them to be hot-dust ULIRGs rather than SMGs.

Furthermore, H11 noted that the observed-frame submm flux, and thus whether a simulated starburst is selected as an SMG or considered a hot-dust ULIRG, depends on the
angle from which the galaxy is viewed. Variation in the SED with viewing angle can cause $\tdust$ to vary for different lines-of-sight, but the variation in $\tdust$ (at most a
few degrees) cannot account for the entire variation in submm flux. Instead, the primary cause is that the effective emission area depends on viewing angle.
Note that in Fig. \ref{fig:SFR-M*} there are many cases where large blue squares and small blue squares are centred at the same ($\mstar$,SFR) point.
These correspond to simulated starbursts that are classified as both SMGs and hot-dust ULIRGs depending on the angle from which they are viewed.
Such cases are relatively common, but there are no quiescently star-forming galaxies for which this occurs; this is most likely because in the quiescently star-forming galaxies
most of the FIR emission arises from regions that are optically thin in the FIR, and thus the effective emitting area in the FIR does not depend strongly on viewing angle.
This interpretation is supported by the results presented in Section \ref{S:obscuration}.

\subsection{Other potential diagnostics} \label{S:other_diagnostics}

This work has focused on diagnostics that rely on integrated UV--mm SEDs alone, but there are various other diagnostics that could potentially
be used to distinguish quiescently star-forming and starburst galaxies. For example, the late-stage merger-induced starburst SMGs may have
higher CO line-widths than the quiescently star-forming disks \citep{Narayanan:2009}. In addition, the CO spectral line energy distributions (SLEDs)
may differ, with the starbursts typically having higher mean excitation than the quiescently star-forming galaxies. Note, however, that the significant dust optical depths present in the starbursts can result in SLEDs that appear indicative of very low excitation conditions, potentially making it
difficult to distinguish highly-obscured high-excitation gas from unobscured low-excitation gas \citep{Papadopoulos:2010b,Papadopoulos:2010}.
X-ray observations may also be useful: In our simulations the merger-induced starbursts typically coincide with strong AGN activity, though
the peak AGN activity occurs $\sim 50-100$ Myr after the peak SFR \citep{Hopkins:2012AGN_delay}. Thus the merger-induced starbursts
should have stronger central X-ray sources than the quiescent discs. Furthermore, since the starbursts are typically more highly obscured,
the AGN they host should have higher hardness ratios than the AGN hosted by quiescently star-forming galaxies. Finally, radio observations
can provide excellent spatial resolution and thus be used to determine the number of components and separation of the nuclei.

\section{Conclusions} \label{S:conclusions}

We have combined high-resolution 3-D hydrodynamic simulations of $z \sim 2$ major mergers of disc galaxies and 3-D Monte Carlo
dust RT calculations to investigate the differences between quiescently star-forming and starburst galaxies.  We have focused on the
SMG population as a case study because, as argued in H11 and elaborated in Section \ref{S:bimodality}, the SMG population is likely a mix of
quiescently star-forming galaxies and merger-induced
starbursts. Our models make robust observational predictions for how quiescently star-forming galaxies
on the so-called `main sequence of star formation' should differ from merger-induced starbursts.
We present multiple observational diagnostics which can
distinguish quiescently star-forming and starburst SMGs based on integrated UV--mm data alone. The testable predictions and observational
diagnostics presented in this work include:

\begin{enumerate}

\item Effective dust temperature -- derived from fitting a single-$T$ OT modified blackbody, the full form of the single-$T$
modified blackbody, or a model assuming a power-law distribution of dust temperatures -- correlates with $\lir$, and the galaxies in the high-$\lir$, high-$\tdust$
region of the $\tdust-\lir$ plot are almost exclusively merger-induced starbursts. A $\tdust$ cut is an effective means to select starbursts
from the overall galaxy population, whereas an $\lir$ cut only works well for those galaxies that are also selected as SMGs.

\item Star formation efficiency, $SFE \equiv \lir/M_{\rm gas}$, is $\sim 2-3 \times$ higher for starburst SMGs than quiescently star-forming SMGs at a given $\lir$.
The starbursts not detected as SMGs have SFE $\sim 2 \times$ greater than the starburst SMGs.

\item The IR excess, IRX $\equiv \lir/\lfuv$, correlates with $\lir$. The starbursts have IRX a factor $\sim 5 \times$
greater than quiescently star-forming galaxies, primarily because they are typically more luminous than quiescently star-forming galaxies
and IRX correlates with $\lir$. At the bright end the correlation is approximately linear, suggesting that $\lfuv$ is decoupled from $\lir$ for
the most obscured starbursts.

\item Effective dust temperature, star formation efficiency, and IRX are all inversely correlated with nuclear separation because the
strength of the merger-driven starburst increases as the galaxies coalesce.

\item The majority of the simulated starbursts lie above the SFR--$\mstar$ relation, whereas the quiescently star-forming galaxies lie close to it.
The observational definition of starbursts as outliers above the relation agrees well with our definition of merger-induced starbursts as simulated
galaxies with instantaneous SFR $>3\times$ their baseline quiescent SFR.

\end{enumerate}

One can apply these observational diagnostics to test our claim that the SMG population is a mix of quiescently star-forming galaxies and merger-induced starbursts
and to constrain the relative contribution of the sub-populations. Furthermore, the tests presented here provide physically
motivated ways to observationally separate quiescently star-forming galaxies and starbursts, enabling one to more cleanly study the underlying physics than
when heterogenous samples (e.g., SMGs) are used.
Though we have focused on the SMG population here, we have also discussed the hot-dust ULIRGs that would not be selected by the SMG selection.
These galaxies tend to be less massive than SMGs. Though they can be as luminous as SMGs, this often requires that they be starbursts, so they
have relatively high $\tdust$ for their $\lir$ and are thus missed by the SMG selection. Hot-dust ULIRGs are also sometimes just SMGs viewed from a different angle,
because the submm flux of a simulated starburst can vary significantly with viewing angle.

We have also explored how well various IR SED fitting forms describe our simulated galaxies' SEDs. For a given SED, the fitted $\tdust$ and $\beta$
can vary significantly depending on what form is used, so these parameters should not be interpreted physically. However, some trends are
robust to choice of fitting method, indicating that the fits still yield useful information about the SED variations amongst galaxies.
In future work we will more thoroughly investigate how well the different
FIR SED fitting methods describe observed and simulated galaxy SEDs, determine how the inferred dust properties relate to the actual physical dust properties,
and develop an improved method for fitting the FIR SEDs of large samples of galaxies.

\acknowledgments
CCH thanks Emanuele Daddi for motivation for this work, Padelis Papadopoulos and Rahul Shetty for lively discussion of IR SED fitting, and Shane Bussmann, Diego Mu\~noz,
and Dominik Riechers for insight into submm interferometry. We thank the anonymous referee, Caitlin Casey, Helmut Dannerbauer, Micha{\l} Micha{\l}owski, Alex Pope, and
Amy Stutz for helpful comments on the manuscript.
We thank Volker Springel for providing the non-public version of \gadgettwo used for this work and Brant Robertson for use of his code to scale the initial conditions to high redshift.
DK is supported by NASA through Hubble Fellowship grant HST-HF-51276.01-A. PJ acknowledges support by a grant from the W. M. Keck Foundation.
The simulations in this paper were performed on the Odyssey cluster supported by the FAS Research Computing Group at Harvard University.
\\

\bibliography{std_citations,smg}

\label{lastpage}

\end{document}